\documentclass[iop, twocolumn]{aastex63}
\hypersetup{citecolor=blue,urlcolor=red}
\shorttitle{Millisecond Pulsars}
\shortauthors{Wang et al.}

\begin{document}

\title{Pulse Jitter and Single-pulse Variability in Millisecond Pulsars}

\correspondingauthor{S.Q. Wang}
\email{wangshuangqiang@xao.ac.cn}

\author{S.Q. Wang}
\affiliation{Xinjiang Astronomical Observatory, Chinese Academy of Sciences, Urumqi, Xinjiang 830011, China}
\affiliation{Key Laboratory of Radio Astronomy, Chinese Academy of Sciences, Urumqi, Xinjiang, 830011, China}
\affiliation{Xinjiang Key Laboratory of Radio Astrophysics, Urumqi, Xinjiang, 830011, People's Republic of China}

\author{N. Wang}
\affiliation{Xinjiang Astronomical Observatory, Chinese Academy of Sciences, Urumqi, Xinjiang 830011, China}
\affiliation{Key Laboratory of Radio Astronomy, Chinese Academy of Sciences, Urumqi, Xinjiang, 830011, China}
\affiliation{Xinjiang Key Laboratory of Radio Astrophysics, Urumqi, Xinjiang, 830011, China}

\author{J.B. Wang}
\affiliation{Institute of Optoelectronic Technology, Lishui University, Lishui, Zhejiang, 323000, China}

\author{G. Hobbs}
\affiliation{CSIRO Astronomy and Space Science, PO Box 76, Epping, NSW 1710, Australia}

\author{H. Xu}
\affiliation{National Astronomical Observatories, Chinese Academy of Sciences, Beijing 100101, China}

\author{B.J. Wang}
\affiliation{National Astronomical Observatories, Chinese Academy of Sciences, Beijing 100101, China}

\author{S. Dai}
\affiliation{School of Science, Western Sydney University, Locked Bag 1797, Penrith South DC, NSW 2751, Australia}

\author{S.J. Dang}
\affiliation{School of Physics and Electronic Science, Guizhou Normal University, Guiyang 550001, China}

\author{D. Li}
\affiliation{National Astronomical Observatories, Chinese Academy of Sciences, Beijing 100101, China}

\author{Y. Feng}
\affiliation{ Zhejiang Lab, Hangzhou, Zhejiang 311121, China}

\author{C.M. Zhang}
\affiliation{National Astronomical Observatories, Chinese Academy of Sciences, Beijing 100101, China}

\begin{abstract}

Understanding the jitter noise resulting from single-pulse phase and shape variations is important for the detection of gravitational waves using pulsar timing array. We presented measurements of jitter noise and single-pulse variability of 12 millisecond pulsars that are part of the International Pulsar Timing Array sample using the Five-hundred-meter Aperture Spherical radio Telescope (FAST). We found that the levels of jitter noise can vary dramatically among pulsars. A moderate correlation with a correlation coefficient of 0.57 between jitter noise and pulse width is detected. 
To mitigate jitter noise, we performed matrix template matching using all four Stokes parameters. Our results revealed a reduction in jitter noise ranging from 6.7\% to 39.6\%.
By performing longitude-resolved fluctuation spectrum analysis, we identified periodic intensity modulations in 10 pulsars. 
In PSR J0030+0451, we detected single-pulses with energies more than 10 times the average pulse energy, suggesting the presence of giant pulses. 
We also observed a periodic mode-changing phenomenon in PSR J0030+0451. 
We examined the achievable timing precision by selecting a sub-set of pulses with a specific range of peak intensity, but no significant improvement in timing precision is achievable.
\end{abstract}


\section{INTRODUCTION}

Pulsar timing enable the most stringent tests of fundamental physics, including the constrain of nuclear physics at extreme densities~\citep{Demorest2010}, testing general relativity~\citep{Kramer2006},  detecting and characterizing low-frequency gravitational waves (GWs, \citealt{Hellings1983,Jenet2005}),  improving the solar system planetary ephemeris~\citep{Guo2019} and developing a pulsar-based timescale~\citep{Hobbs2020}.  
By monitoring the pulse times of arrival (ToAs) of an ensemble of the most stable millisecond pulsars (MSPs), known as the pulsar timing array (PTA), it is possible to detect nanohertz GWs~\citep{Detweiler1979,Hellings1983,Foster1990,Jenet2005}. 
The nanohertz GW sources include inspiraling supermassive black hole binaries, decaying cosmic string networks, relic post-inflation GWs, and even non-GW imprints of axionic dark matter~\citep{Burke-Spolaor2019}. 

There are currently six individual PTAs, namely the European Pulsar Timing Array (EPTA; \citealt{Kramer2013}), the North American Nanohertz Observatory for Gravitational Waves (NANOGrav; \citealt{McLaughlin2013}), the Parkes Pulsar Timing Array (PPTA; \citealt{Manchester2013}), the MeerKAT Pulsar Timing Array (MPTA; \citealt{Miles2023}, the Indian Pulsar Timing Array (InPTA; \citealt{Tarafdar2022}), and the Chinese Pulsar Timing Array (CPTA; \citealt{Xu2023}). 
The timing data of thses individual PTAs have been used to search for GWs, e.g. \citet{Agazie2023, Antoniadis2023, Reardon2023, Xu2023}.
The individual PTAs also collaborate under the International Pulsar Timing Array (IPTA), pooling their data sets to improve the detection sensitivity of GWs~\citep{Hobbs2010,Manchester2013,Verbiest2016,Perera2019}.

The success of GW detection with PTA requires the highest timing precision possible. 
The precision of ToAs is limited by many noise contributions, including those introduced by the pulsar itself, the interstellar medium along the line of sight, and the measurement process~\citep{Cordes2010, Lam2016,Verbiest2018}.
On short timescales, the excess noise in the timing residuals is typical dominated by white noise, which includes radiometer noise,  jitter noise and scintillation noise~\citep{Liu2012,Shannon2014,Lam2019}. 
Radiometer noise describes the noise temperature/power in the profile, and its magnitude depends on the signal-to-noise ratio (S/N) of the average profile. 
Jitter noise is a stochastic process that affects all pulsars and is induced by intrinsic variations in single-pulse amplitude and phase~\citep{Cordes1985}. 
Scintillation noise is caused by changes in the interstellar impulse response from multipath scattering~\citep{Cordes2010}. 
The pulse broadening function results from diffractive interstellar scattering/scintillation (DISS), which depends on observational frequency strongly.

Jitter noise is expected to dominate the white noise budget in high S/N observations~\citep{Shannon2014}. 
Several studies have been conducted to investigate the effects of jitter noise on pulsar timing.  
Using Parkes telescope, \citet{Oslowski2011} determined that the timing precision of PSR J0437$-$4715 is limited to approximately 30\,ns in an hour duration due to the presence of jitter noise.
By measuring the jitter noise of 22 MSPs as a part of PPTA, \citet{Shannon2014} found that  PSR J1909-3744 shows the lowest levels of jitter noise of about 10\,ns in an hour duration. \citet{Lam2019} detected jitter noises in 43 MSPs as part of NANOGrav and found that the level of jitter noise is frequency-dependent. 
Recently, \citet{Parthasarathy2021} conducted jitter measurements in 29 pulsars using the MeerKAT radio telescope and identified that PSR J2241$-$5236 shows the lowest jitter level of about 4\,ns in an hour duration.

The Five-hundred-meter Aperture Spherical radio Telescope (FAST) is the most sensitive single dish telescope, providing an opportunity to detect and constrain jitter noises in MSPs, e.g. \citet{Feng2021,Wang2020,Wang2021}. 
In the paper, we present measurements of jitter noise and single-pulse variability in 12 MSPs using FAST. 
In Section 2, we describe our observations.
In Section 3, we present the measurements of jitter noises, the single-pulse properties, and examine the achievable timing precision by using a subset of pulses with a specific range of peak intensity.  
we discuss and summarize our results in Section 4.

\section{OBSERVATIONS AND DATA PROCESSING}

For our observations of 12 MSPs, we used the central beam of the 19-beam receiver of FAST with a frequency range of 1000 to 1500\,MHz~\citep{Jiang2019}. 
Each pulsar was observed once, the Modified Julian Date (MJD) and duration of each pulsar are shown in Table~\ref{tab}. Data for the pulsars were recorded in search mode PSRFITS format with four polarizations, 8-bit samples at an interval of 8.192\,$\mu s$, and 1024 frequency channels with a channel bandwidth of 0.488\,MHz. Note that FAST does not possess the signal processing capability to provide coherently dedispersed search mode data streams.

For each pulsar, individual pulses were extracted with ephemerides from the Australia Telescope National Facility (ATNF) pulsar catalogue (\citealt{Manchester05}), using the {\sc dspsr} software package~\citep{Straten2011}. To remove radio frequency interference (RFI), we used {\sc paz} in the {\sc psrchive} software package~\citep{Hotan2004} to zap channels using median smoothed difference, and a degraded bandpass of about 50\,MHz was removed from both edges of the data. Then, we used {\sc pazi} to check RFI for each pulsar by eye. Polarization calibration was achieved by correcting for the differential gain and phase between the receptors through separate measurements using a noise diode signal.

Rotation Measure (RM) is determined using {\sc rmfit}, and the results are presented in Table~\ref{tab}. Subsequently, the pulses for each pulsar are RM-corrected. To generate ToAs, noise-free standard templates were created using the {\sc psrchive} program {\sc paas}, and ToAs were subsequently calculated by cross-correlating the pulse profile with the template using {\sc pat} with the FDM algorithm. Timing residuals were obtained using the {\sc tempo2} software package~\citep{Hobbs2006}. Fluctuation analysis was conducted using the PSRSALSA package~\citep{Weltevrede2016}.

\section{Results}

\subsection{Jitter noise measurement}

\begin{table*}  
\centering
\tiny
\caption{Jitter measurements for 12 MSPs. Columns 2, 3, 4 and 5 present the period, DM, observed MJD, and integration time, respectively, columns 6 and 7 present the measured RM and maximum $S/N_{\rm peak}$ of a single-pulse, respectively, columns 8 and 9 present the implied jitter value which scales to single-pulse and one-hour-long duration, respectively, column 10 and 11 present the implied $\sigma_{\rm S/N}$ and $\sigma_{\rm DISS}$ in one-hour-long duration, respectively. The last column reports a reference to a previously published jitter measurement at L-band, in which P21, L19, L16 and S14 refer to \citet{Parthasarathy2021} at 1284\,MHz,  \citet{Lam2019} at 1500\,MHz, \citet{Lam2016} at 1400\,MHz and \citet{Shannon2014} at 1400\,MHz, respectively. } 
\label{tab}
\begin{tabular}{ccccccccccccc}
\hline
NAME &  Period &  DM & MJD &  Duration  & RM & ${\rm S/N_{\rm peak,max}}$ & $\sigma_{\rm J}$(1) & $\sigma_{\rm J}$(h) & $\sigma_{\rm S/N}$(h) & $\sigma_{\rm DISS}$(h) & $\sigma_{\rm J}$(h) 
\\
 &  ms & ${\rm cm^{-3}\,pc}$    &  &  s & $\rm rad\,m^{-2}$ & & $\mu$s &ns  & ns & ns  & ns
\\
\hline
{ J0030+0451} & 4.86    & 4.34 & 59454.80& 570 &   2.44$\pm$0.21 &     69.93   & 40.1$\pm$0.8&46$\pm$1 &  43.4$\pm$0.1 &0.27 &  <60 (P21)
\\
    &      &   & &    &   &  &  &  &     &   &  $61.1_{-3.5}^{+3.5}$ (L19) 
    \\
  &      &   &  &  &   &  &    &    &  &   &  153 (L16)
\\
{ J0613$-$0200}& 3.06  &  38.78  & 59478.03& 571  &  21.79$\pm$0.23  & 10.47&  37.3$\pm$0.2 &  34.4$\pm$0.2&  33.0$\pm$0.1 &6.62 &$133_{-8}^{+8}$ (L19)
 \\
   &      &   &    &   &  &  &  &    &   &   &  <44 (L16) 
        \\
  &      &   &    &   &  &   & &  &   &   &  <400 (S14) 
\\
{ J0636+5128} & 2.86  & 11.11 &59132.92 &558&   -2.14$\pm$0.05  & 16.89 &33.2$\pm$0.4 & 29.6$\pm$0.4 & 12.33$\pm$0.07 &2.48 &  $62_{-22}^{+16}$ (L19)
\\
{ J0751+1807} & 3.48  & 30.25 & 59478.04 & 421 & 42.53$\pm$0.19   &  12.71  &33.1$\pm$0.5  & 32.5$\pm$0.5& 29.26$\pm$0.08 &6.00 & - 
\\
{J1012+5307}&  5.26  & 9.02 &59446.18& 535 & 3.84$\pm$0.05 &   16.99 &  22.3$\pm$0.1 &27.0$\pm$0.2 &  29.20$\pm$ 0.06 &1.98 & $67\pm 6$ (L19)
        \\
  &      &   &    & &  &  &    &   &&   &   <103 (L16) 
\\
{J1643$-$1224} & 4.62  &  62.41 & 59479.37 &  421 &  -296.71$\pm$0.23   &9.25 & 36.0$\pm$0.2 &40.8$\pm$0.3& 39.89$\pm$0.03 &12.06 &  <60 (P21)
 \\
   &      &   &    &   &  &    &   & &    &   &  $120_{-10}^{+8}$ (L19)
\\
    &      &   &    &   &  &    &  & &  &   &  155 (L16)
        \\
  &      &   &    &   &  &    &   & &  &   &  <500 (S14) 
\\
{J1713+0747}  & 4.57   & 15.92 & 59446.46&575 & 13.16$\pm$0.08   & 62.18 &   22.1$\pm$0.2 & 24.9$\pm$0.3& 6.43$\pm$0.01 &3.65 & $29.3_{-0.9}^{+3.2}$ (L19)
        \\
  &      &   &    &   &  &    &  &    & &  &  28/36 (L16)
          \\
  &      &   &    &   &  &    &   &   & &  &  $35.0\pm 0.8$ (S14)
\\
{ J1744$-$1134} &  4.08 &  3.14   & 59486.39&  397 &  5.77$\pm$0.09 &11.53  &  27.6 $\pm$ 0.2  & 29.4 $\pm$ 0.2 & 28.4$\pm$0.1 &0.11 &  30$\pm$6 (P21)
 \\
   &      &   &    &   &  &    &  &   & &  &  46.5$\pm$1.3 (L19)
\\
    &      &   &    &   &  &    &  &   &  & &  31 (L16)
        \\
  &      &   & &   &   &  &    &  & &  &  37.8$\pm$ 0.8 (S14)
\\
{J1911+1347} &  4.63 & 30.99  &  59477.61&  571  &  -7.85$\pm$0.31 & 18.58&   24.5$\pm$0.2 & 27.8$\pm$0.2  &  27.7$\pm$0.1 &4.47 & $43.5_{-2.3}^{+2.3}$ (L19)
\\
{J1918$-$0642}  & 7.65  & 26.46 & 59449.54& 1169 & -56.24$\pm$0.19   & 33.73   & 38.2$\pm$0.3 & 55.7$\pm$0.5&  41.2$\pm$0.2 &5.35 & <55 (P21)
\\
    &      &   &    &   &  &    &  & &  &   & $59_{-13}^{+9}$ (L19) 
        \\
  &      &   &    &   &  &    &  &  &  &   & <101 (L16) 
\\
{J1944+0907} & 5.19  & 24.36  &59449.51 & 1169 &  -34.61$\pm$0.17  &29.37 &   145$\pm$1& 174$\pm$1&  79$\pm$1 &3.54 & $311_{-7}^{+6}$ (L19)
        \\
  &      &   &    &   &  &    &  &  &  &   &  196 (L16) 
\\
{J2145$-$0750}  &  16.05 & 9.00 & 59463.65& 1171 &  0.42$\pm$0.05 &132.28 &  79.5$\pm$0.4 & 168$\pm$1 &22.6$\pm$0.1 &  1.37 & 200 $\pm$ 20  (P21)
\\
  &  & &  & &  & &   &  & &  & $173 ^{+3}_{-4} $ (L19)
\\
  &  & &  & &  & &   &  & &  & 85   (L16)
\\
  &  & &  & &  & &  & &  & & 192 $\pm$ 6  (S14)
\\ 
\hline
 \end{tabular}
\end{table*}

\begin{figure*}
\centering
\includegraphics[width=130mm]{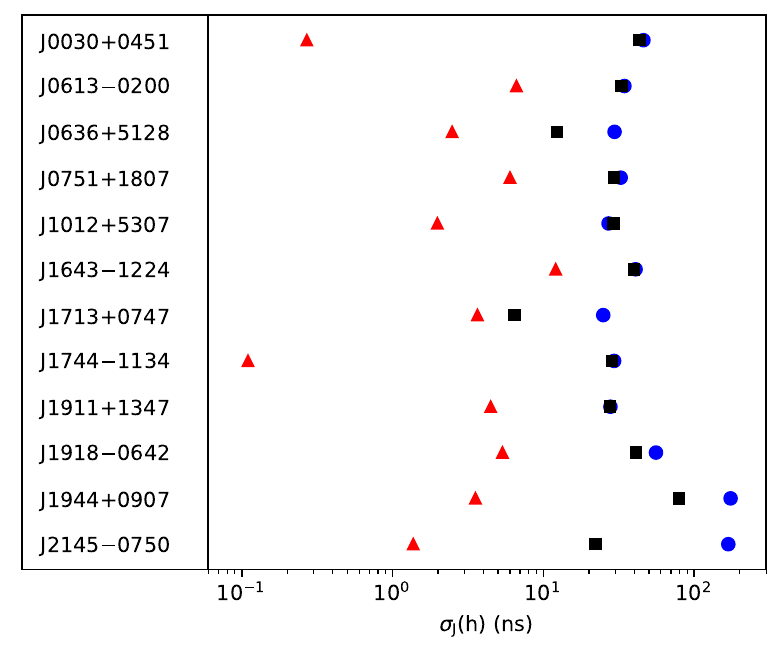}
\caption{Summary of white noise components for 12 pulsars. The right panel shows the three contributions, radiometer noise as black squares, jitter noise as blue circles, and scintillation noise as red triangles. All the noise contributions are scaled to one hour integration.}
\label{noise}
\end{figure*}

On short timescales, the excess noise in the timing residuals is typical dominated by radiometer noise, jitter noise, and scintillation noise~\citep{Shannon2014, Lam2016}. 
The measurement error of ToA can be summarized as~\citep{Shannon2014,Lam2016}: 
\begin{equation}\label{eq}
\sigma^2_{\rm total}=\sigma^2_{\rm S/N}+\sigma^2_{\rm J}+\sigma^2_{\rm DISS},
\end{equation}
where $\sigma_{\rm S/N}$, $\sigma_{\rm J}$ and $\sigma_{\rm DISS}$ are the uncertainties induced by radiometer noise, jitter noise and scintillation noise, respectively. 

We assume that all of the excess error in the ToA measurements is attributed to jitter noise. 
Jitter noise can then be obtained by calculating the quadrature difference between the observed root mean square (rms) timing residual and the radiometer noise~\citep{Shannon2014}: 
\begin{equation}\label{eq}
\sigma_{\rm J}^2{(N_{\rm p})}=\sigma_{\rm obs}^2{(N_{\rm p})}-\sigma_{\rm sim}^2{(N_{\rm p})},
\end{equation}
where ${N_{\rm p}}$ is the number of averaged pulses, $\sigma_{\rm obs}^2$ and $\sigma_{\rm sim}^2$ are the variances of the observed ToAs and simulated ToAs, respectively. 
Using the method that was presented in~\citet{Parthasarathy2021}, we simulated a set of idealized TOAs which match the model, then Gaussian noise is added to the idealized TOAs with magnitudes equal to the uncertainties of the observed TOAs. Using the TEMPO2 FAKE tool~\citep{Hobbs2006}, we created 1000 realizations of these simulated data sets for each pulsar, and the mean $\sigma_{\rm sim}^2$ is taken as the variance of the  simulated ToAs.  
Following \citet{Parthasarathy2021}, we think that the jitter noise is detected if $\sigma_{\rm obs}^2{(N_{\rm p})} > 0.95\sigma_{\rm sim}^2{(N_{\rm p})}$. 

For each pulsar,  the profiles are  frequency-averaged as well as polarization-averaged, and then we folded the profile with ${N_{\rm p}}$ of 16, 32, 64, 128, 256, 512, 1024, 2048, and 4096 single pulses. 
Subsequent, the jitter noises of these 12 pulsars were measured using frequency-averaged ToAs. 
Note that we used only the total intensity (Stokes I) of pulses to generate the frequency-averaged ToAs. 
Considering the jitter noises generally can be scaled as $1/\sqrt{N_{\rm p}}$ (e.g. \citealt{Shannon2014}), the extrapolated jitter noise for a one-hour-long duration is calculated as: $\sigma_{\rm J}({\rm h})=\sigma_{\rm J}(1)/\sqrt{3600/P}$, where $\sigma_{\rm J}({1})$ is the implied jitter noise for a single-pulse, and $P$ is the pulsar spin period in seconds. The extrapolated jitter noise levels for single-pulse and one-hour-long duration are presented in Table~\ref{tab}. 

In our sample, we have detected jitter noises in all 12 pulsars, with different levels of jitter noise for each pulsar.
The fraction of jitter noise ($F_{\rm J}=\sigma_{\rm J}$/$\sigma_{\rm total}$) ranges from 68\% to 99\%. Jitter noise is particularly dominant in bright pulsars, such as PSR J2145$-$0750 with an $F_{\rm J}$ of 99\%.
A jitter parameter is defined as $k_{\rm J}=\sigma_{\rm J}(1)/P$~\citep{Lam2019}, where $P$ is the pulsar spin period. We analyzed the relationship between $k_{\rm J}$ and the duty cycle for each pulsar, where the duty cycle is defined as the full width at half maximum of the pulse profile (W50) divided by the pulse period ($P$).
We found a Spearman correlation coefficient of $R=0.82$ between jitter noise and pulse W50 width for the pulsars in our sample. However, we note that PSR J1944+0907 exhibits both high jitter noise and a wide pulse profile, which may introduce biases in the estimation of the correlation coefficient. When PSR J1944+0907 is excluded, the correlation coefficient decreases to 0.56. Therefore, we suggested that there is no significant relationship between the level of pulse jitter and the pulse width. Our result is consistent with previous studies conducted by NANOGrav~\citep{Lam2019} and MPTA~\citep{Parthasarathy2021}, which reported correlation coefficients of 0.62 and 0.64, respectively.

\begin{figure*}
\centering
\includegraphics[width=40mm]{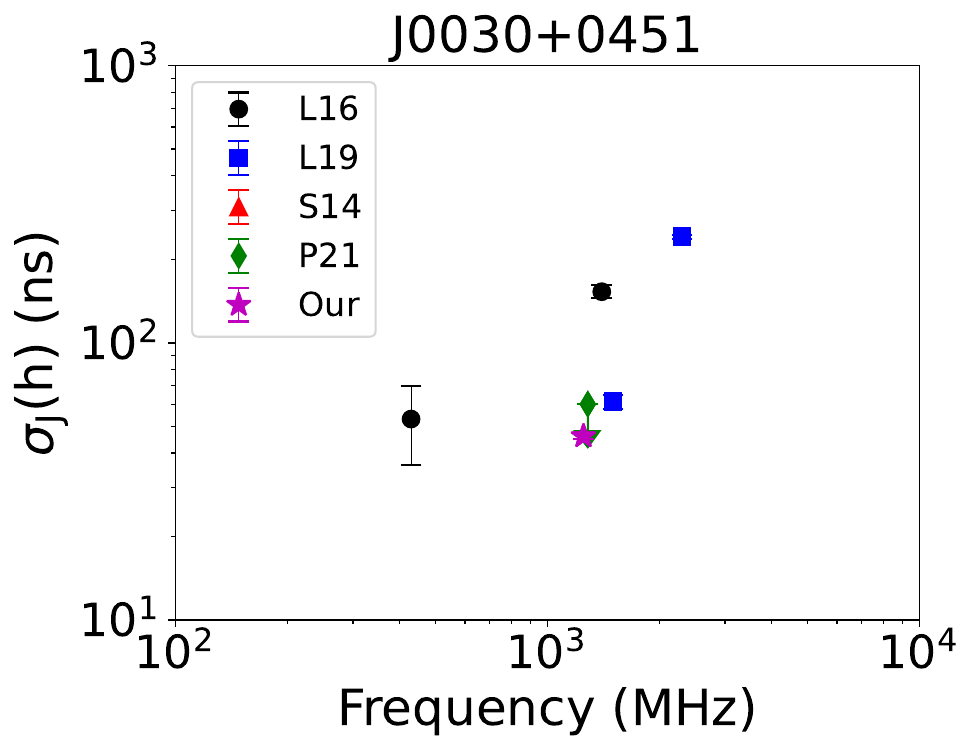}
\includegraphics[width=40mm]{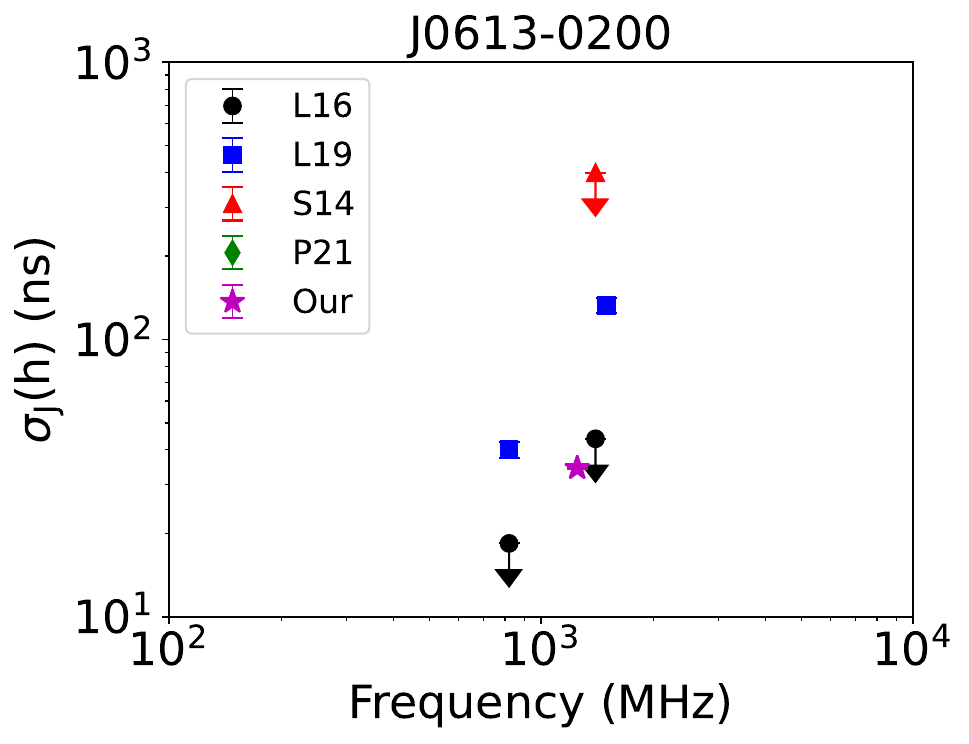}
\includegraphics[width=40mm]{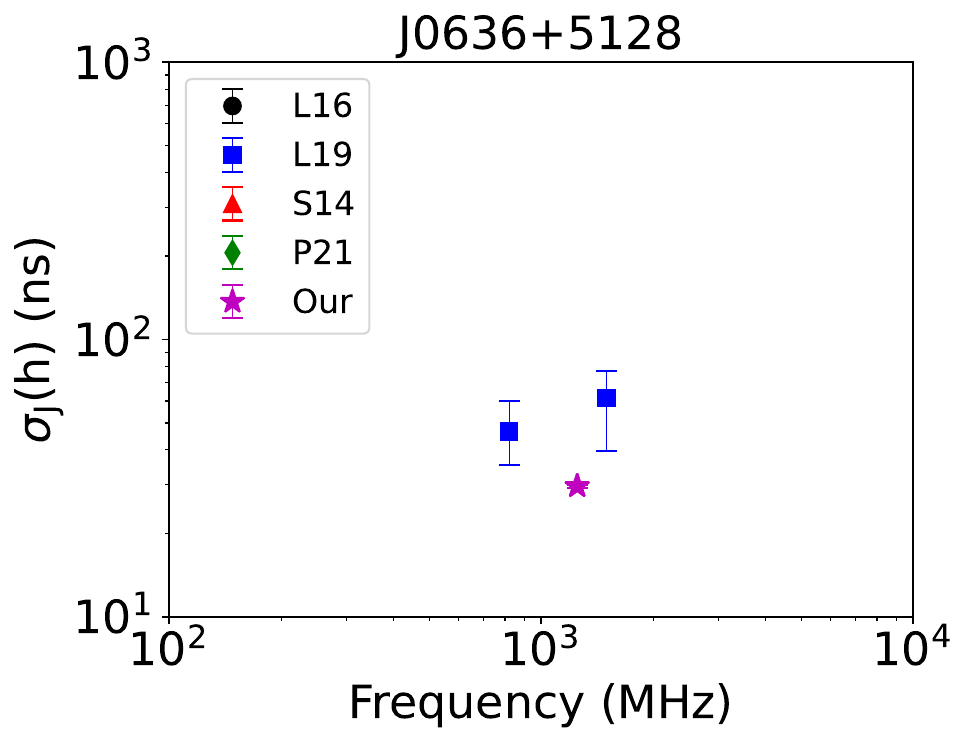}
\includegraphics[width=40mm]{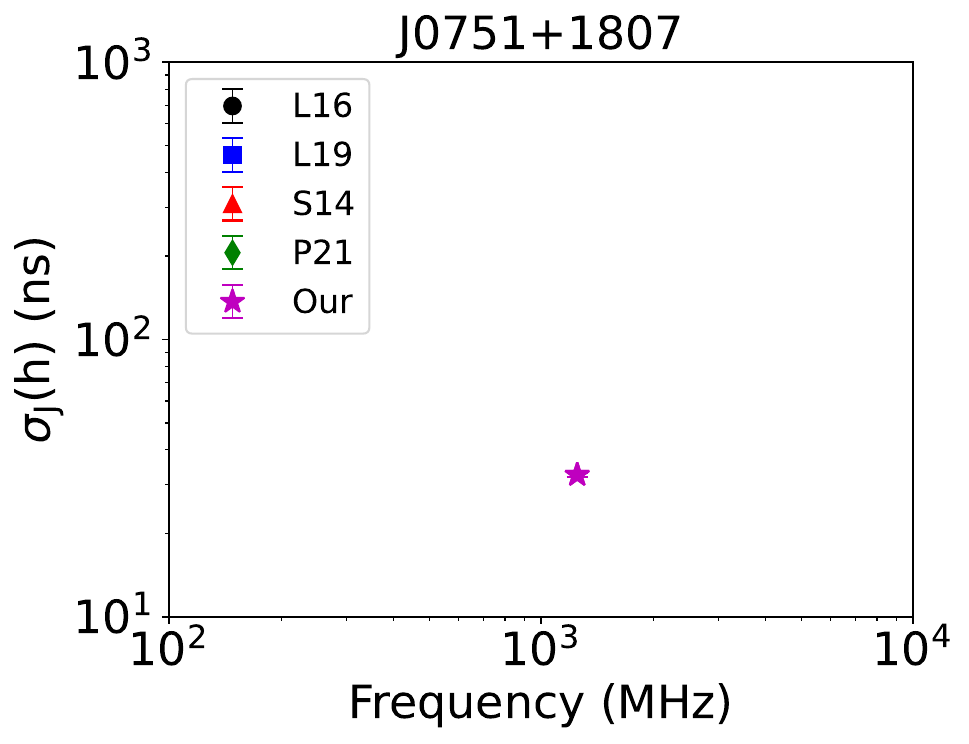 }
\includegraphics[width=40mm]{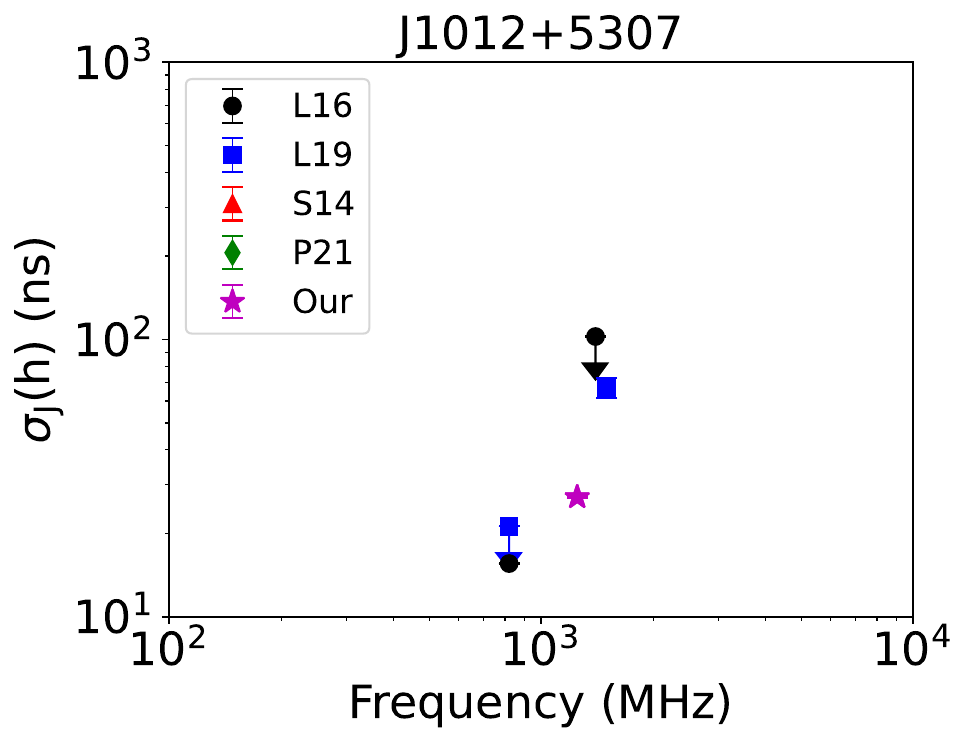}
\includegraphics[width=40mm]{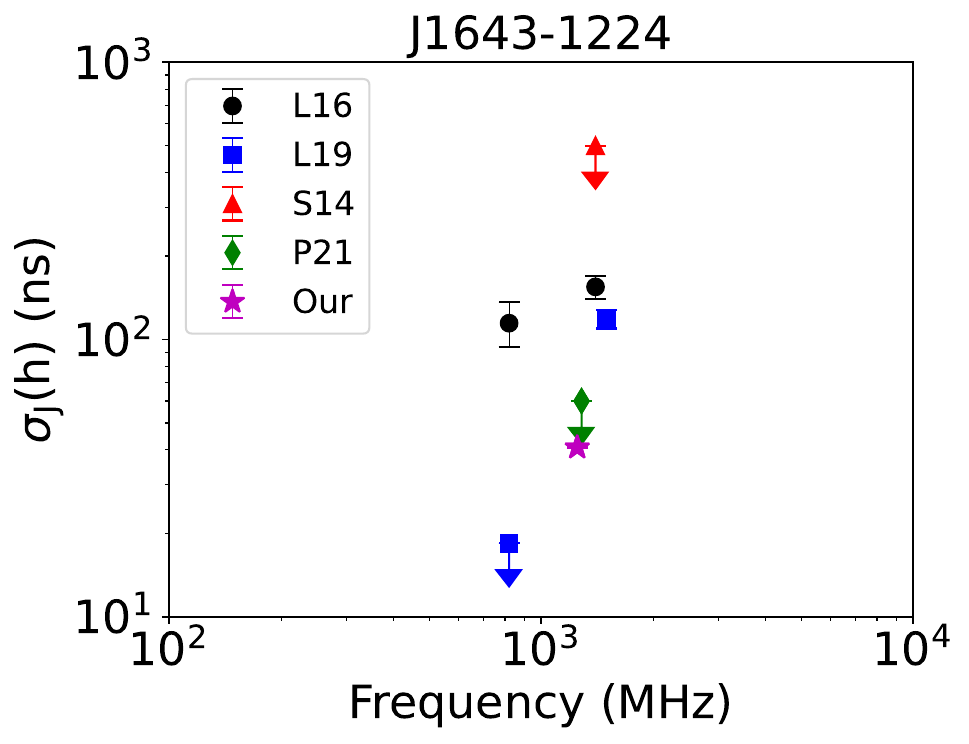}
\includegraphics[width=40mm]{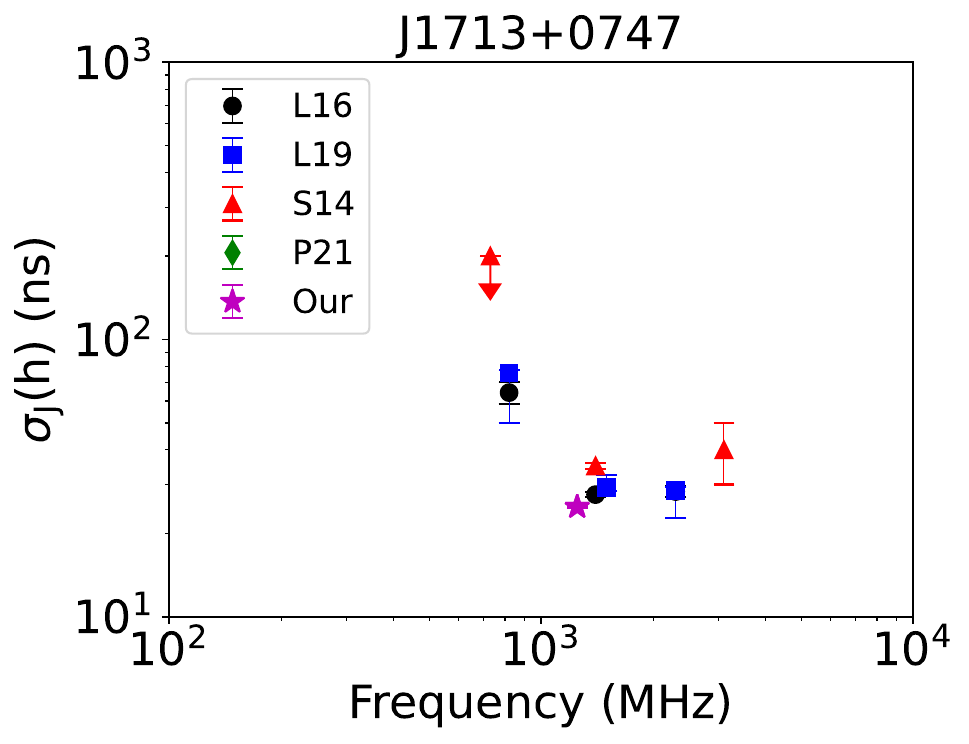}
\includegraphics[width=40mm]{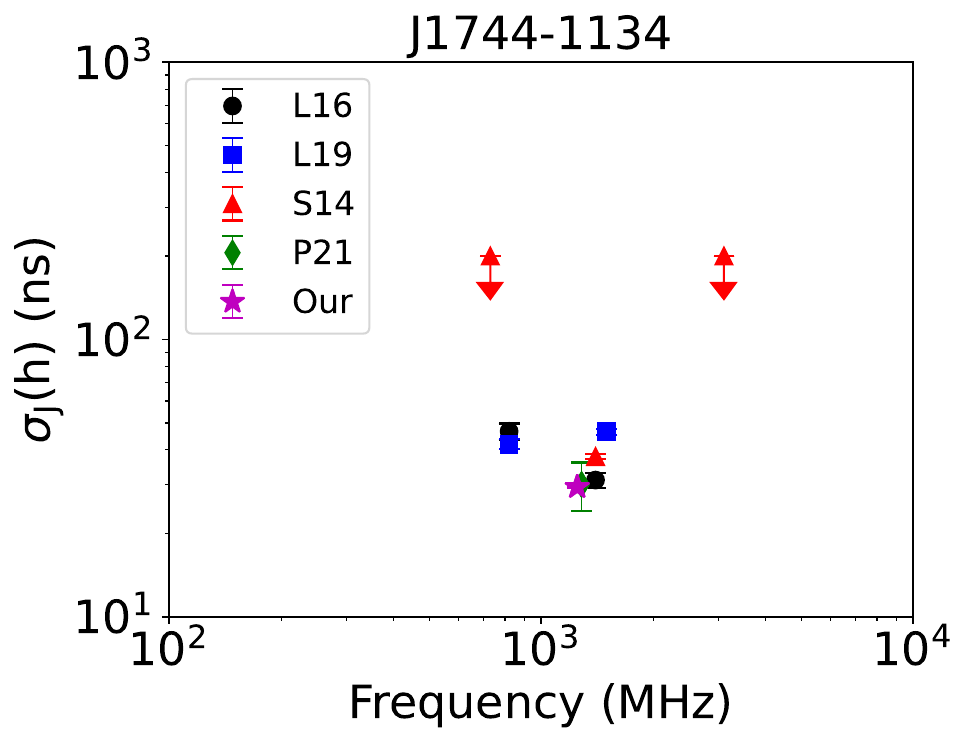}
\includegraphics[width=40mm]{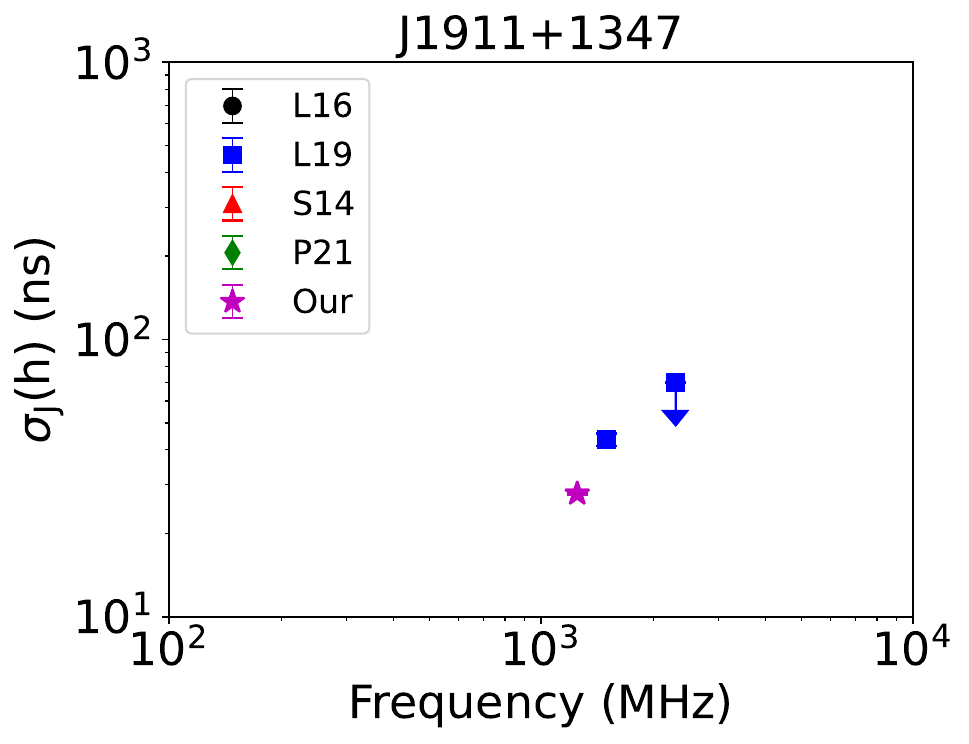}
\includegraphics[width=40mm]{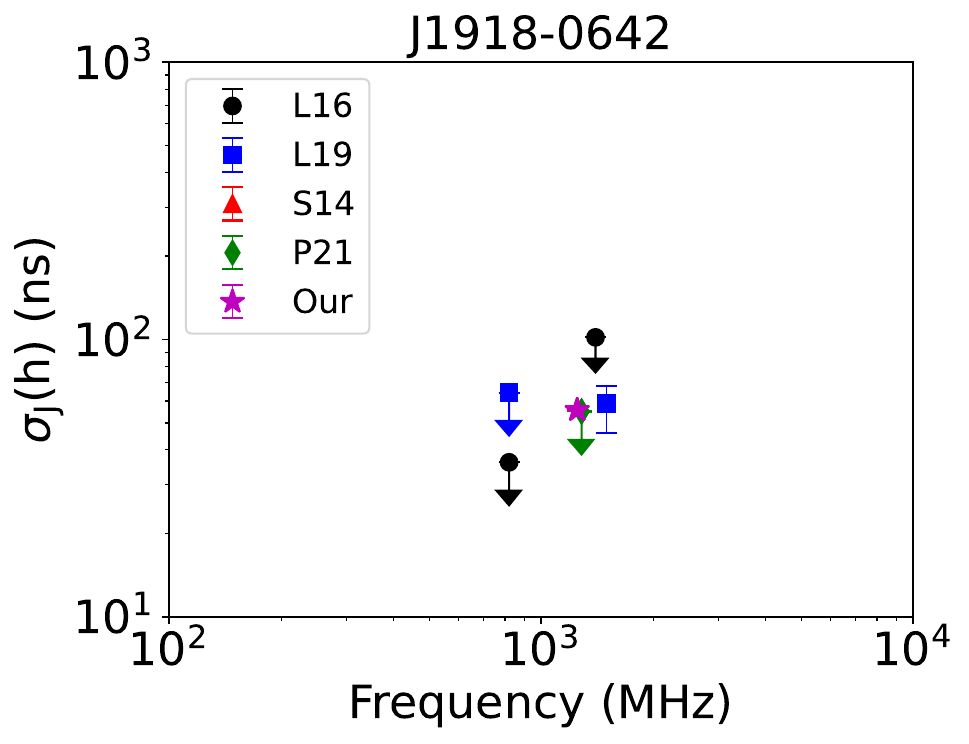}
\includegraphics[width=40mm]{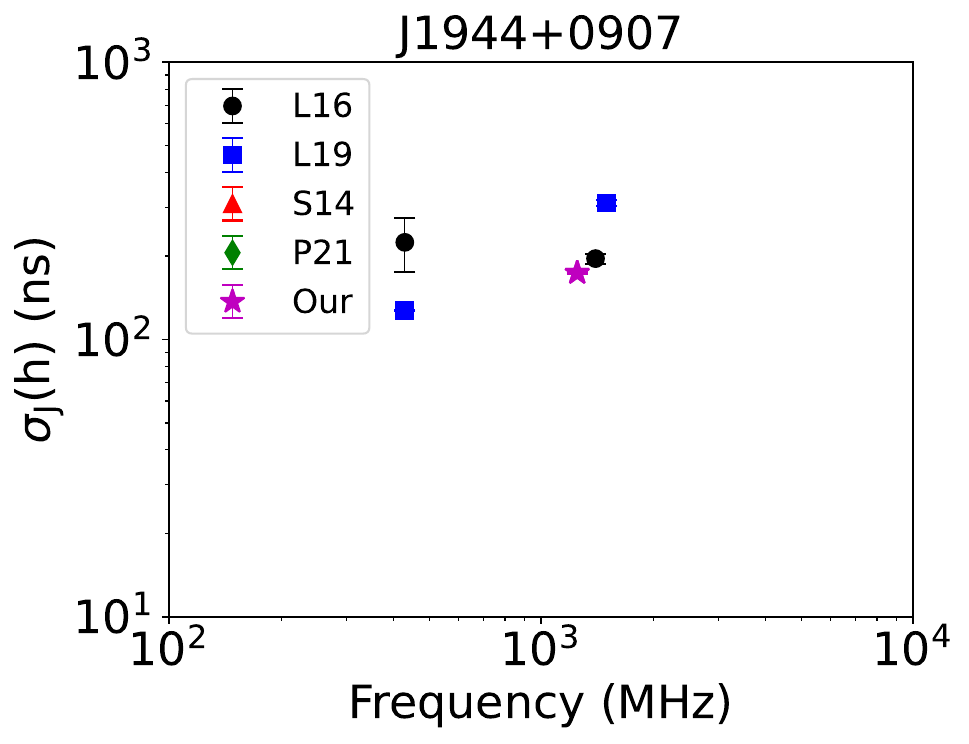}
\includegraphics[width=40mm]{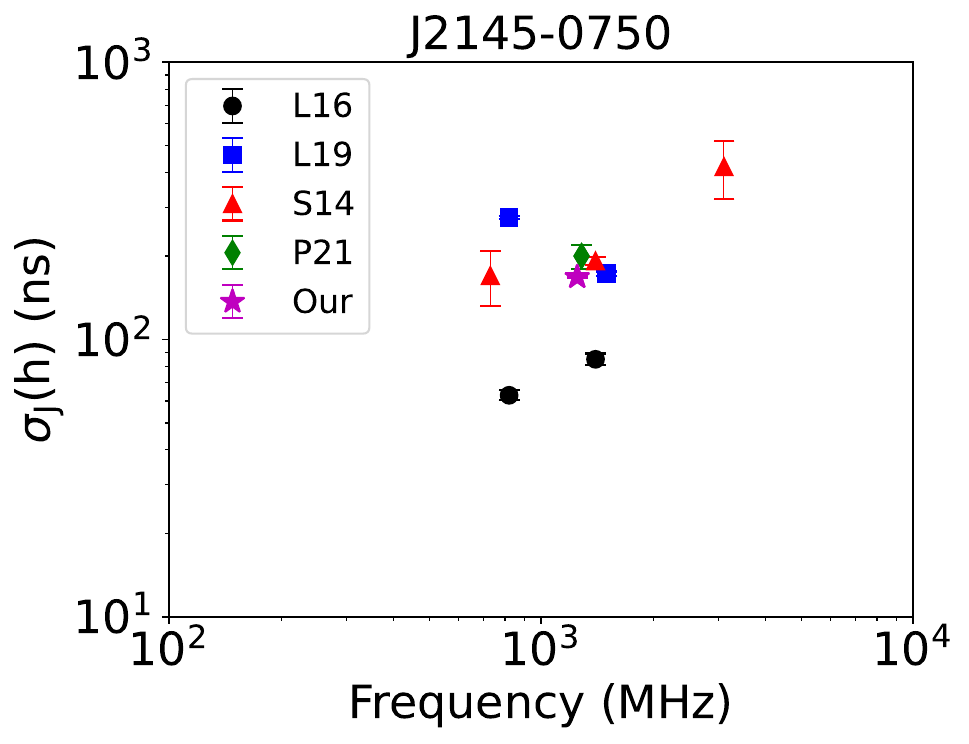}
\caption{The jitter noises at different frequencies for 12 pulsars. The black circles denote the measured jitter noise at  430\,MHz, 820\,MHz, 1400\,MHz and 2300\,MHz of \citet{Lam2016}, the blue squares denote the measured jitter noise at  327/430\,MHz, 820\,MHz, 1500\,MHz and 2300\,MHz of \citet{Lam2019}, the red triangles denote the measured jitter noise at 730\,MHz, 1400\,MHz and 3100\,MHz of \citet{Shannon2014}, the green diamond denote the measured jitter noise at 1284\,MHz of \citet{Parthasarathy2021}, and the magenta stars denote the measured jitter noise at 1250\,MHz of our work. 
Note the levels of jitter noise are scaled to one-hour-long duration.}
\label{fj}
\end{figure*}

In Figure~\ref{fj}, we showed our results and the previously reported values of jitter noise at different frequencies for each pulsar. Jitter noise for many pulsars exhibits frequency dependence, with different levels at different observing frequencies~\citep{Lam2019,Parthasarathy2021}. 
We compared our measurements of jitter noise with previously reported values. Details of the comparison for each pulsar are provided in Section 3.3.

Scintillation noise is associated with the propagation of pulsar signals through the interstellar medium. 
The scintillation noise which is induced by stochastic broadening can be described as~\citep{Cordes2010}
\begin{equation}\label{eq}
\sigma_{\rm DISS}^2=\tau ^2/N_{\rm scint},
\end{equation}
where $\tau$ and $N_{\rm scint}$ are the pulse-broadening time-scale  and the number of scintles, respectively.
The number of scintles  $N_{\rm scint}=(1+\eta \Delta \nu/\nu_{\rm d})(1+\eta \Delta T/ t_{\rm d})$, where $\Delta \nu$ and $\Delta T$ are the observing bandwidth and duration, $\nu_{\rm d}$ and $t_{\rm d}$ are the diffractive scintillation bandwidth and time, respectively. 
The scintillation filling factor is approximately $\eta \approx 0.3$~\citep{Cordes2010}.
Using the NE2001 model~\citep{Cordes2002}, we estimated $\nu_{\rm d}$, $t_{\rm d}$, and $\tau$ for each pulsar. Taking $\Delta \nu=500$\,MHz and $\Delta T=3600\,$s, the $\sigma_{\rm DISS}$ in one-hour-long duration for each pulsar is estimated.

In Figure~\ref{noise}, we present a summary of radiometer noise (black squares), jitter noise (blue circles), and scintillation noise (red triangles) for the 12 pulsars. It is evident that $\sigma_{\rm scint}$ is typically much smaller than $\sigma_{\rm J}$ or $\sigma_{\rm S/N}$ in our sample. The mean levels of jitter noise, radiometer noise, and scintillation noise in one-hour-long duration are 58\,ns, 33\,ns, and 4\,ns, respectively.

\subsection{Reduce jitter noise using polarization}

Our measurements of white noise components in pulsars indicate the substantial impact of jitter noise on PTA with FAST.
For the generation of ToAs, narrow features in the pulse profile can provide strong constraints during template matching~\citep{Straten06}. 
Considering that the polarized components of pulsar profile may exhibit sharp features (e.g. \citealt{Dai2015,Wahl2022}), we tried to use polarization information to mitigate jitter noise. 
We performed matrix template matching using all four Stokes parameters (\citealt{Straten06,Oslowski13}) to generate frequency-averaged ToAs for each pulsar. 
Then, using the same jitter measurement method as shown in Section 3.1, we measured the radiometer noise ($\sigma_{\rm S/N, P}$) and jitter noise ($\sigma_{\rm J, P}$) for each pulsar. 
However, we only detected jitter noises in 8 pulsars. 
The extrapolated noise levels for a one-hour-long duration for these 8 pulsars are shown in Table~\ref{tabp}.

\begin{figure}
\centering
\includegraphics[width=70mm]{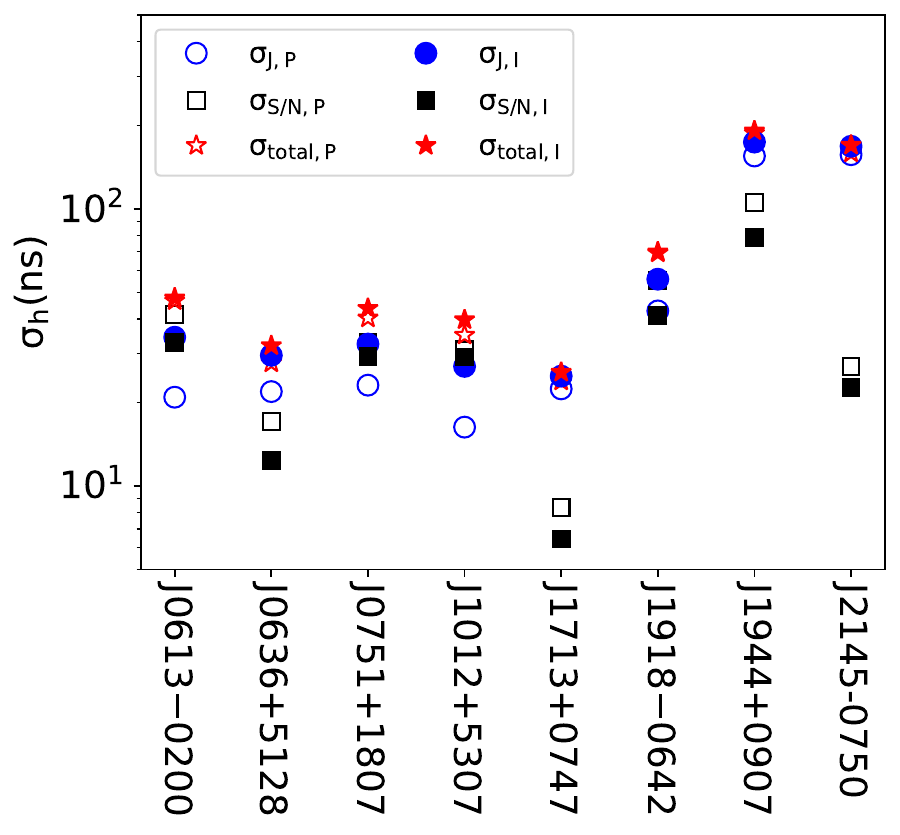}
\caption{Comparison of white noise components for 8 pulsars by using all four Stokes parameters (open symbols) and only I (filled symbols), respectively. The circle, square and star are for the jitter, radiometer and total white noises, respectively. }
\label{p}
\end{figure}

In Figure~\ref{p}, we compared the jitter and radiometer noises which were measured using all four Stokes parameters with that measured using only the total intensity.
The level of radiometer noise increases, which is expected because the Stokes Q, U, and V are generally much weaker than the Stokes I. Simultaneously, the level of jitter noise decreases.
For these 8 pulsars, the radiometer noise level increases by percentages ranging from 6.1\% to 38.6\%, while the jitter noise level decreases by percentages ranging from 6.7\% to 39.6\%.
Therefore, matrix template matching using all four Stokes parameters proves to be a valuable method for reducing jitter noise in pulsars. 
In the case of 7 pulsars, the total white noise level decreases by percentages ranging from 0.19\% to 13.4\%. However, for J1918$-$0642, it increases by 0.08\%. 
The consideration of all four Stokes parameters could lead to improved timing precision on short timescales for most pulsars (7 out of 8 pulsars in our sample).

\begin{table}  
\centering
\scriptsize
\caption{The measured jitter, radiometer and total white noises in one-hour-long duration for 8 pulsars, the subscripts P and I are for the results using all four Stokes parameters and only I, respectively.} 
\label{tabp}
\begin{tabular}{ccccccccccccc}
\hline
NAME &  $\sigma_{\rm J, P}$(h)       &    $\sigma_{\rm S/N, P}$(h)    &  $\sigma_{\rm total, P}$(h)  &   $\sigma_{\rm total, I}$(h) 
\\
 & ns  &  ns   & ns & ns  
\\
\hline
{ J0613$-$0200}& 20.9$\pm$0.5  &  41.63$\pm$0.05 &  46.8 &  47.7
\\
{ J0636+5128} &   21.9$\pm$0.4  &   17.09$\pm$0.06   & 27.8 &  32.1
\\
{ J0751+1807} &     23.1$\pm$0.5   &    33.09$\pm$0.09   & 40.4 &  43.7
\\
{J1012+5307}&     16.3$\pm$0.2   &    30.98$\pm$0.03    & 35.0 &  39.8
\\
{J1713+0747}  &    22.4$\pm$0.3   &    8.37$\pm$0.02   &  23.9 &  25.7
\\
{J1918$-$0642}  &    42.8$\pm$0.9   &    55.2$\pm$0.2    & 69.8 &  69.3
\\
{J1944+0907} &    155$\pm$1   &     105.4$\pm$0.2  &  187.4 &  191.1
\\
{J2145$-$0750}  &     156.7$\pm$0.7   &    26.92$\pm$0.08   &  159.0 &  169.5
\\
\hline
 \end{tabular}
\end{table}

\subsection{Single-pulse phenomenology}

\begin{figure*}
\centering
\includegraphics[width=40mm]{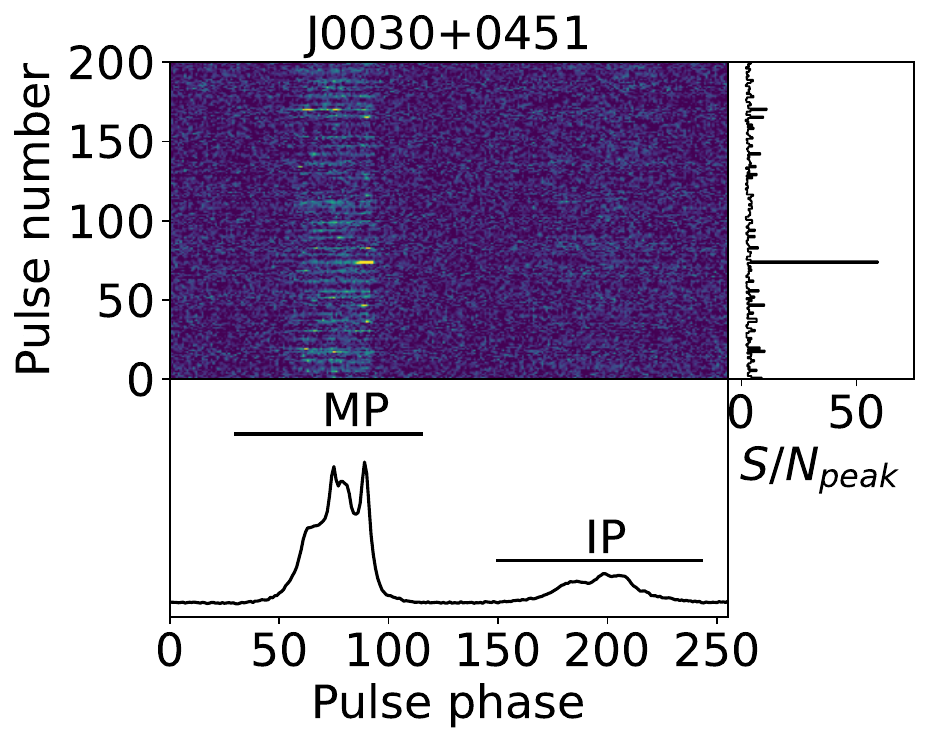}
\includegraphics[width=40mm]{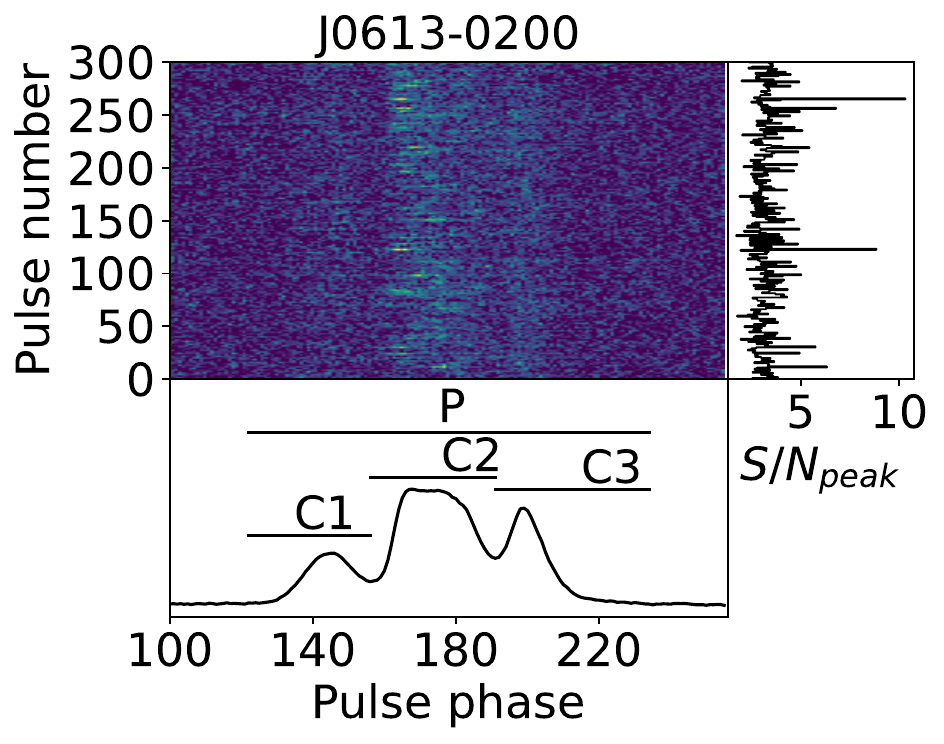}
\includegraphics[width=40mm]{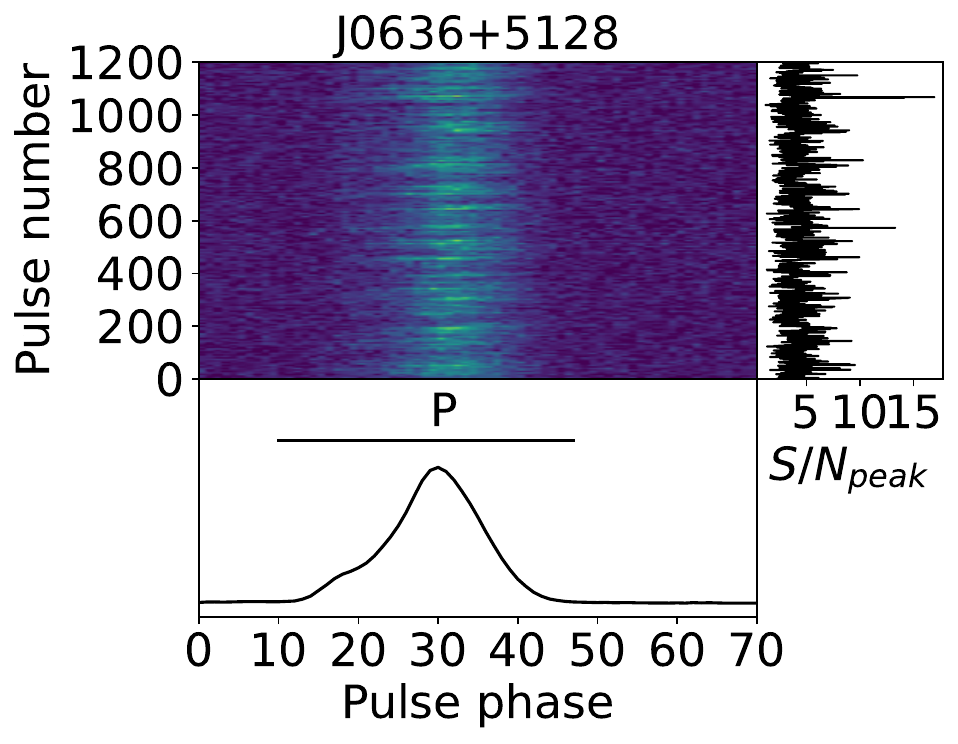}
\includegraphics[width=40mm]{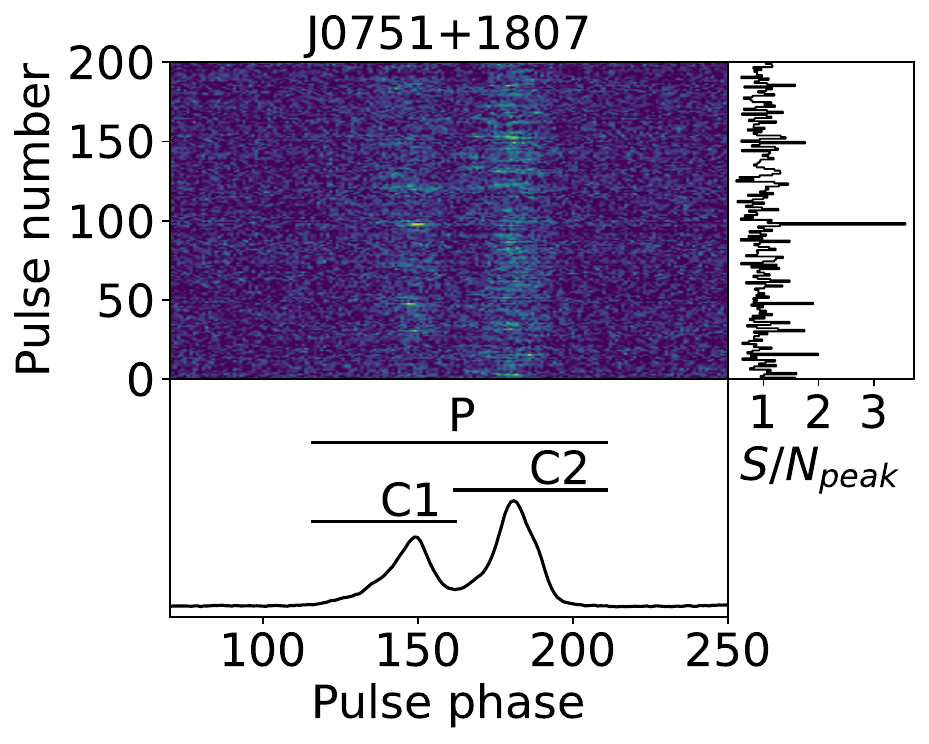}
\includegraphics[width=40mm]{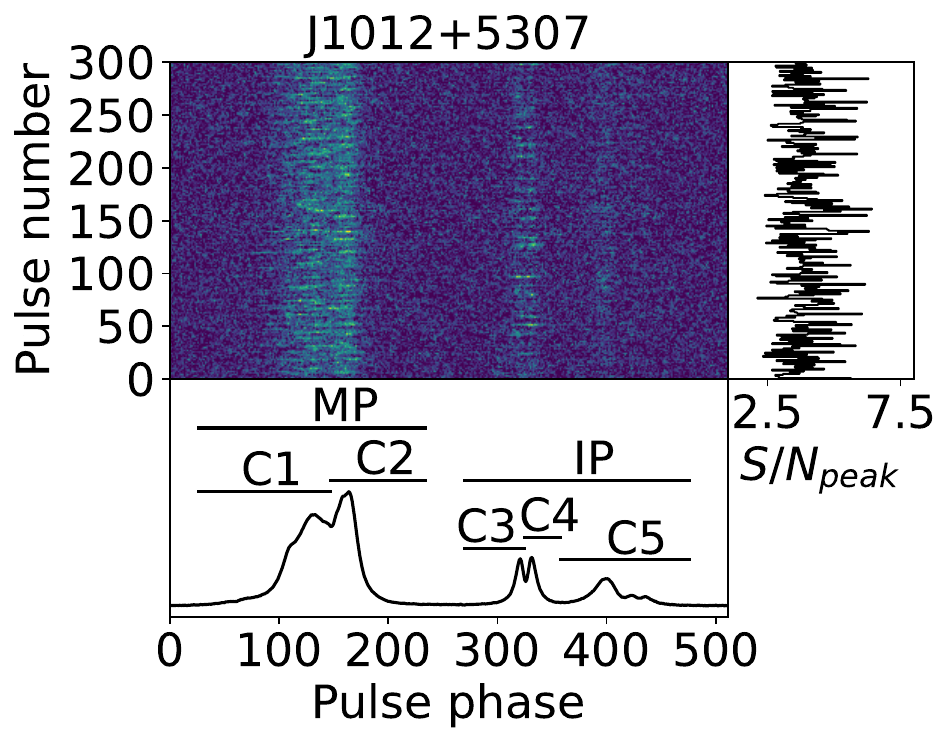}
\includegraphics[width=40mm]{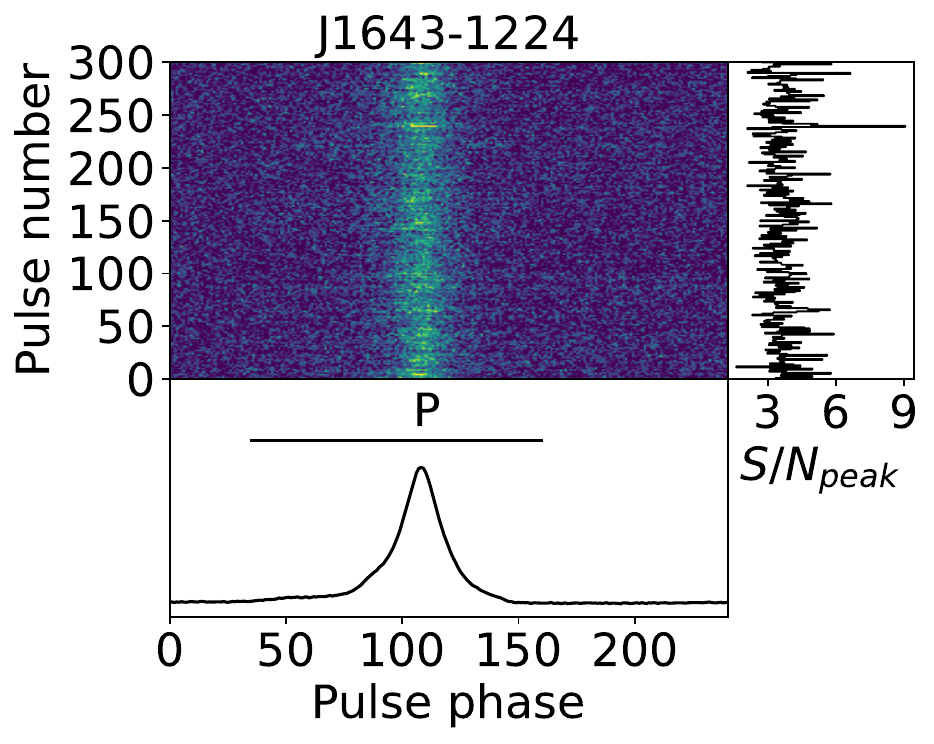}
\includegraphics[width=40mm]{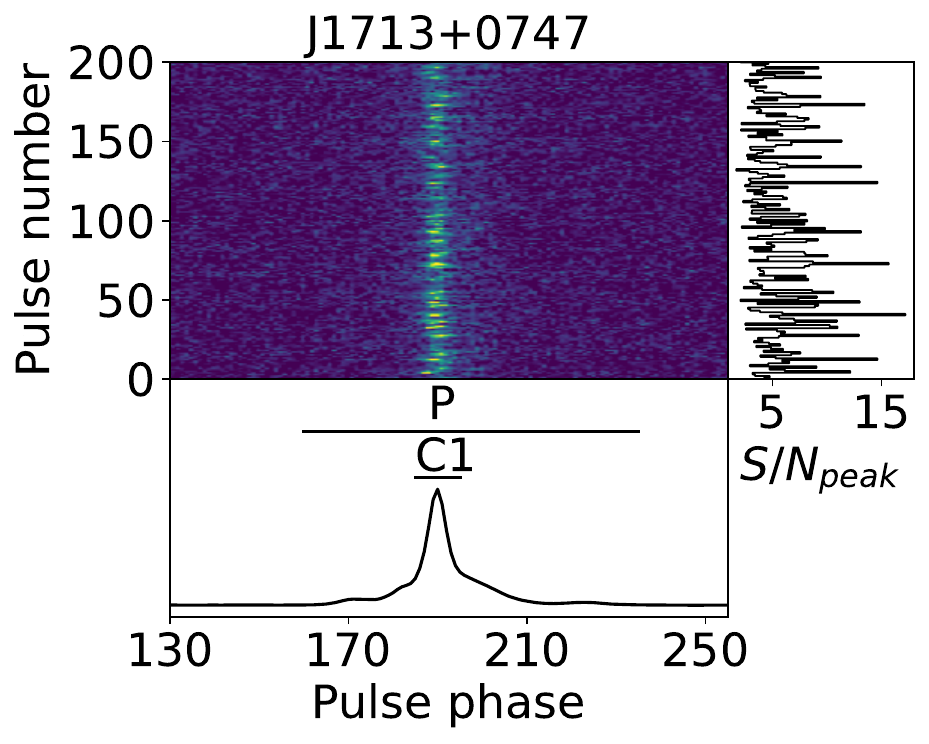}
\includegraphics[width=40mm]{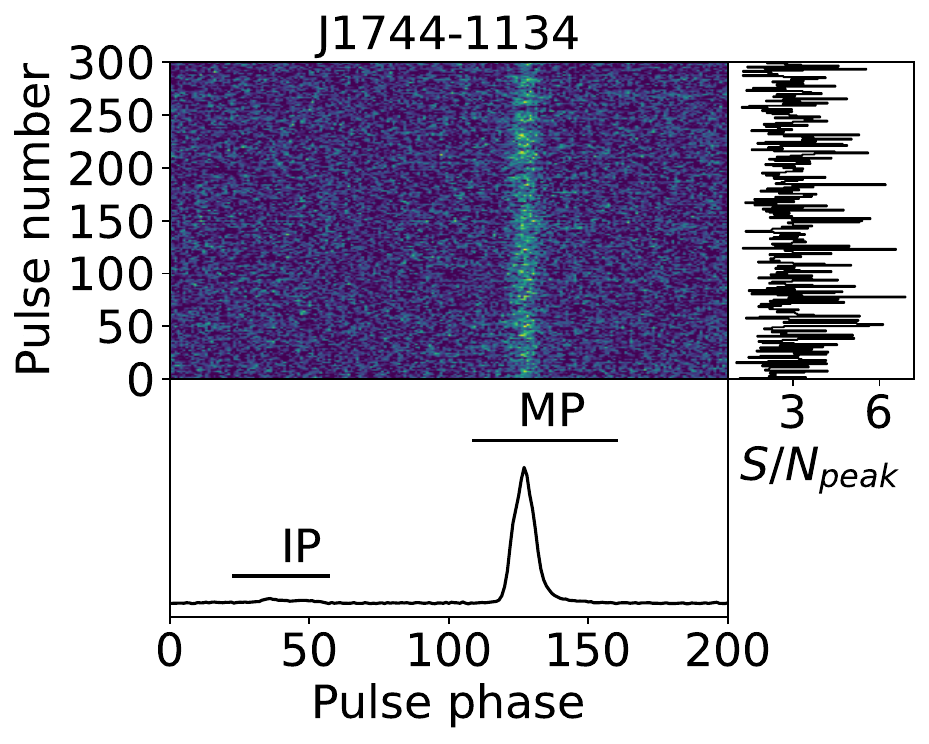}
\includegraphics[width=40mm]{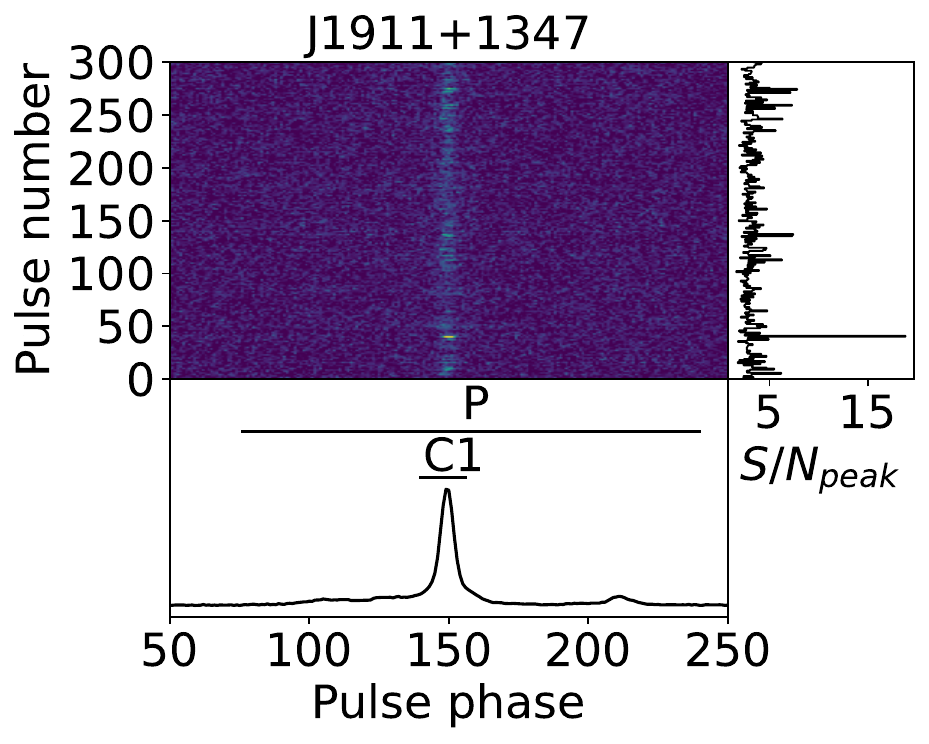}
\includegraphics[width=40mm]{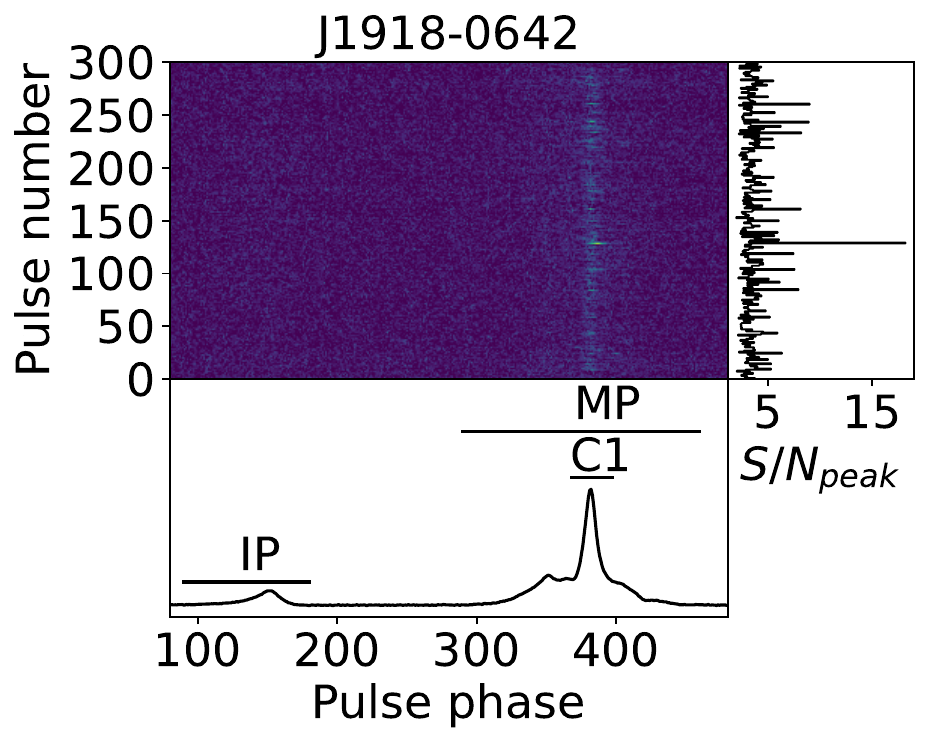}
\includegraphics[width=40mm]{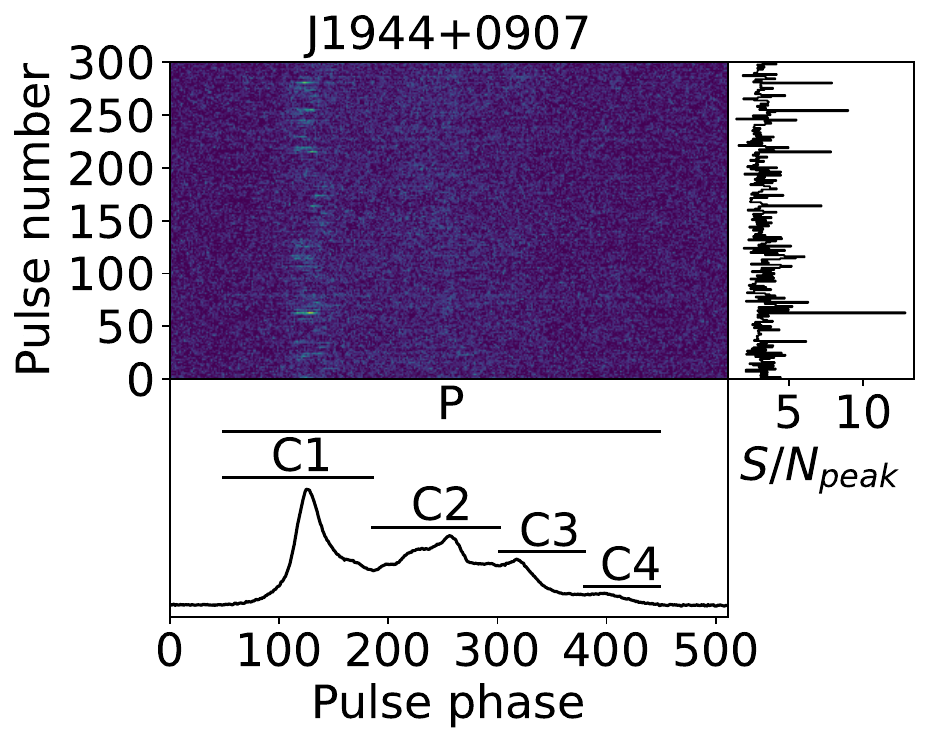}
\includegraphics[width=40mm]{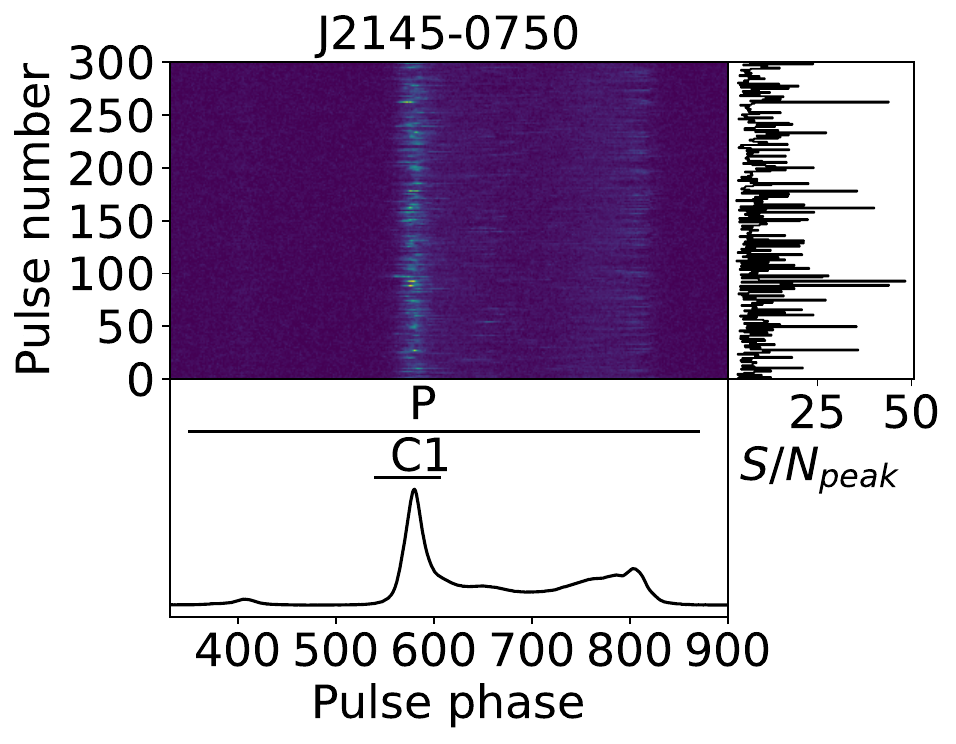}
\caption{Single-pulse stacks for each pulsar. The $S/N_{\rm peak}$ of single-pulse versus pulse number is shown in the left panel and the average profile formed from all of single-pulses is shown in the lower panel, in which different pulse components are labled. }
\label{gray}
\end{figure*}

\begin{figure*}
\centering
\includegraphics[width=40mm]{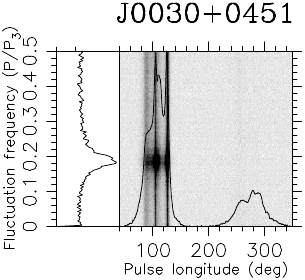}
\includegraphics[width=40mm]{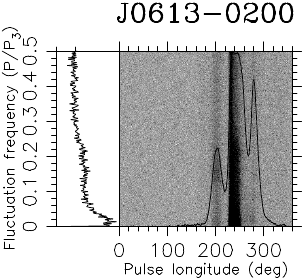}
\includegraphics[width=40mm]{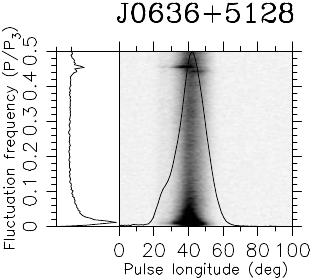}
\includegraphics[width=40mm]{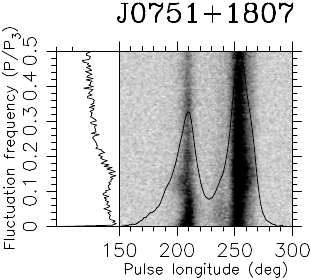}
\includegraphics[width=40mm]{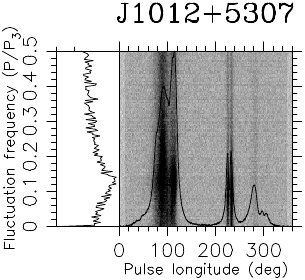}
\includegraphics[width=40mm]{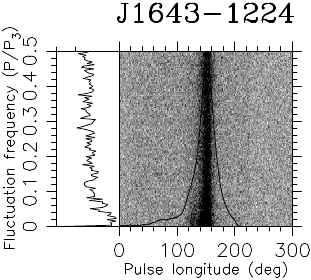}
\includegraphics[width=40mm]{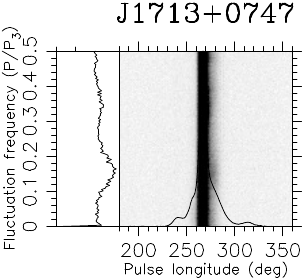}
\includegraphics[width=40mm]{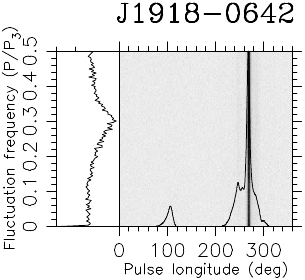}
\includegraphics[width=40mm]{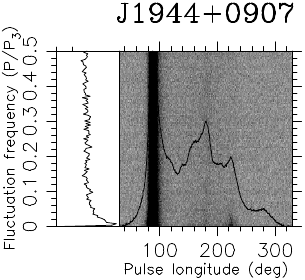}
\includegraphics[width=40mm]{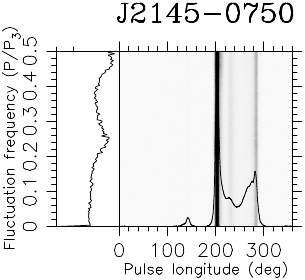}
\caption{The LRFS for 10 pulsars. The slide panels show the horizontally integrated power of LRFS. The averaged profile is overlayed over LRFS. }
\label{modulation}
\end{figure*}

\begin{table}  
\centering
\footnotesize
\caption{Pulse intensity modulation periods in 10 pulsars.} 
\label{p3}
\begin{tabular}{ccccccccc}
\hline
PSR &  Component & Modulation period (P) & Type
\\
\hline
{J0030+0451}    & MP &  5.5$\pm$0.1 &  coherent
\\
{J0613$-$0200}  & C1 &  6$\pm$1.   &  diffuse
\\
                            & C2 & 72.9$\pm$ 0.4   &  diffuse
\\
{J0636+5128}   & P &  84.3$\pm$ 0.9   &  coherent
\\
                          & LP &  2.20$\pm$ 0.01 &  coherent
\\
                          & TP &  2.20 $\pm$  0.01 &  coherent
\\
{J0751+1807}   & MP &  128.0 $\pm$ 0.1/9.4 $\pm$ 0.1 &  diffuse
\\
                         & MP &   6.3 $\pm$ 0.1 &  diffuse
\\
{J1012+5307}   & C1 & 7.1$\pm$ 0.1  &  diffuse
\\
                         & C2 & 7.1$\pm$ 0.1  &  diffuse
\\
                         & C3/C4 & 11.1$\pm$ 0.1  &  diffuse
\\
{J1643$-$1224}   & P & 46$\pm$1   &  diffuse
\\
{J1713+0747}   & C1 &  6.2$\pm$0.1 &  diffuse
\\
{J1918$-$0642}   & C1 & 3.3 $\pm$ 0.1  &  diffuse
\\
{J1944+0907}   & C1/C2/C3 & 128 $\pm$ 2  &  coherent
\\
{J2145$-$0750}   & C1 & 3.9$\pm$ 0.1/ 2.0$\pm$ 0.1 &  diffuse
\\

\hline
 \end{tabular}
\end{table}

We present the single-pulse stacks for these 12 pulsars in Figure~\ref{gray}.
It is worth noting that there are some bright pulses with narrow pulse widths.
Therefore, we define the $S/N_{\rm peak}$ of a single-pulse as $I_{\rm peak}/\sigma_{\rm off}$, where $I_{\rm peak}$ represents the peak intensity of a single-pulse in the on-pulse region and $\sigma_{\rm off}$ is the root-mean-square (rms) of the off-pulse region. 
The $S/N_{\rm peak}$ of single-pulses versus pulse number for each pulsar is shown in the right panel of Figure~\ref{gray}, and the average profile of the entire observation is presented in the lower panel Figure~\ref{gray}.
We found that the single-pulses in MSPs are unstable and exhibit significant variations in $S/N_{\rm peak}$.

To investigate pulse intensity modulation behavior, we conducted a longitude-resolved fluctuation spectrum (LRFS, \citealt{Backer1970,Weltevrede2016}) analysis. Using the {\sc PSRSALSA} package~\citep{Weltevrede2016}, we obtained the LRFS for each pulsar, and we observed that 10 pulsars exhibit periodic intensity modulations, as shown in Figure~\ref{modulation}. In the side panel of Figure~\ref{modulation}, the spectra of horizontally integrated LRFS are presented, where the vertical axis is measured in cycles per period (${P/P_3}$) with the vertical band separation ${P_3}$. The peak of the side panel corresponds to the modulation period of pulse intensity.
Following the method of \citet{Weltevrede2016}, if the vertical extension of ${P_3}$ is smaller than 0.05 cycles per period, it is defined as coherent modulation; otherwise, it belongs to diffuse modulation.

The modulation periods of the pulse intensities for pulsars are shown in Table~\ref{p3}. There are 10 pulsars exhibit periodic intensity modulations, namely PSR J0030+0451, PSR J0613$-$0200, PSR J0636+5128, PSR J0751+1807, PSR J1012+5307, PSR J1643$-$1224, PSR J1713+0747, PSR J1918$-$0642, PSR J1944+0907, and PSR J2145$-$0750, with 6 pulsars being reported for the first time. The periodic intensity modulations of PSR J1918$-$0642, PSR J1012+5307, PSR J2145$-$0750, and PSR J1713+0747 were previously reported by \citet{Edwards2003} and \citet{Liu2016}, respectively, which are consistent with our results.
The comparisons for these four pulsars to the previously published results are shown in the follow subsections.
 PSR J0030+0451, PSR J0636+5128, and PSR J1944+0907 exhibit coherent modulations, while the remaining 7 pulsars show diffuse modulations. The pulse intensity modulation periods of these pulsars range from $\sim 6-665$\,ms ($\sim 2-128\,P$ ).

\subsubsection{PSR J0030+0451}

\begin{figure}
\centering
\includegraphics[width=70mm]{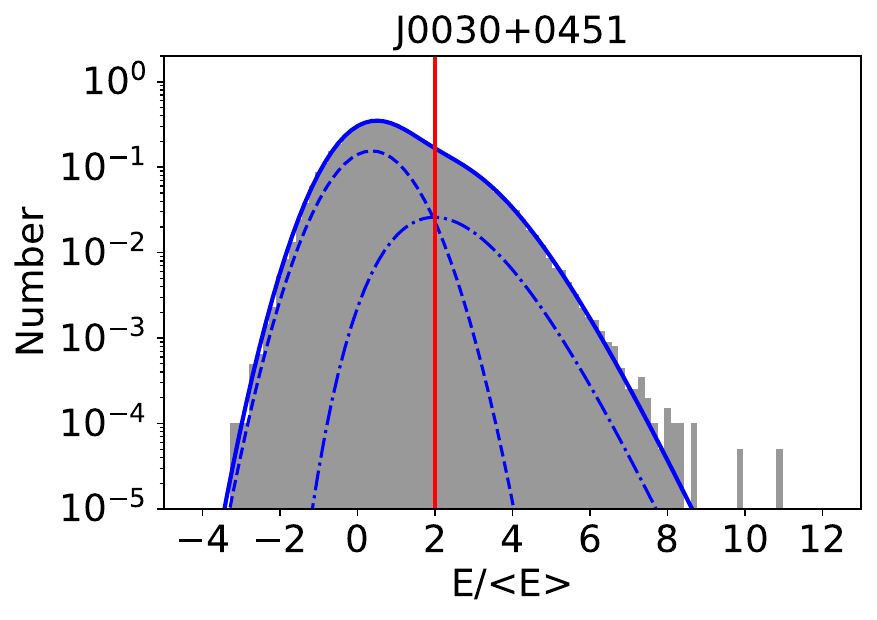}
\caption{Energy distribution for the on-pulse region of PSR J0030+0451. The energies are normalized by the mean on-pulse energy. The blue solid line represents the fitting results of the combination of a Gaussian (blue dashed line, represents the weak mode) and a log-normal (blue dot-dashed line, represents the bright mode) components. The vertical red line represents the 2$\langle E\rangle$ where the Gaussian and log-normal functions intersect. }
\label{0030energy}
\end{figure}

\begin{figure}
\centering
\includegraphics[width=70mm]{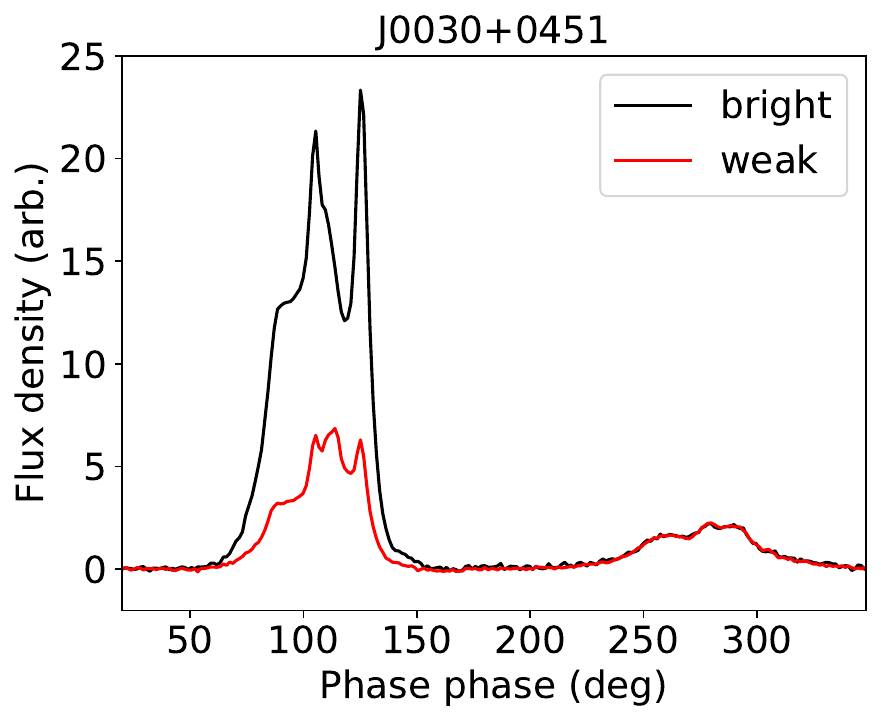}
\caption{Averaged profile of the bright (black line) and weak (red line) modes for PSR J0030+0451. }
\label{0030mode}
\end{figure}

\begin{figure}
\centering
\includegraphics[width=70mm]{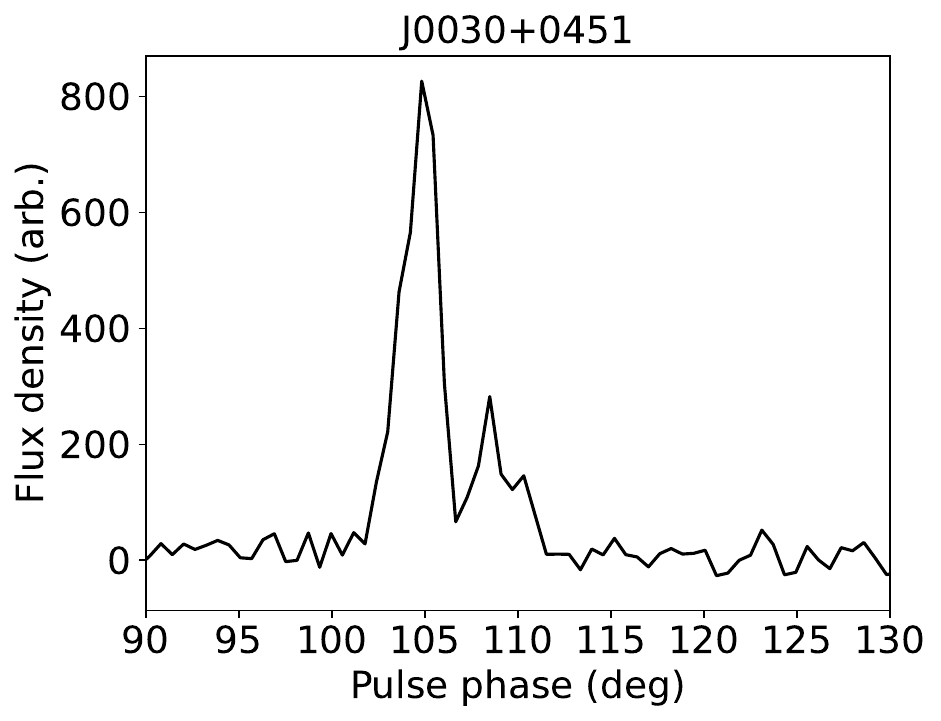}
\caption{Profile of the brightest single-pulse with the pulse energy of 11$\langle E\rangle$ for PSR J0030+0451. }
\label{0030sp}
\end{figure}

PSR J0030+0451 exhibits a main pulse (MP) and a relatively weak interpulse (IP) (Figure~\ref{gray}). 
The energy distribution of the MP is shown in Figure~\ref{0030energy}, where pulse energy is calculated across the on-pulse region labeled as MP in Figure~\ref{gray}. 
This distribution exhibits two peaks, which can be fitted by combination of a Gaussian and a log-normal functions, corresponding to the weak and bright modes, respectively. The Gaussian and log-normal functions intersect at approximately 2$\langle E\rangle$ (marked by the vertical red line in Figure~\ref{0030energy}). To identify the two modes, we employ a criterion where a single pulse with energy greater than 2$\langle E\rangle$ is classified as belonging to the bright mode, otherwise, it is assigned to the weak mode.

In Figure~\ref{0030mode}, the averaged profiles of the weak and bright modes are presented. In the bright mode, the profile exhibits a significantly stronger main pulse compared to the weak mode, while the interpulses have the same shape and flux density in both modes. LRFS analysis reveals that the MP shows periodic intensity modulations with a period of 5.5$\pm0.1\,P$, whereas no such periodic modulations are observed in the interpulse (Figure~\ref{modulation}). The periodical modulation spans the entire profile of the MP. We concluded that PSR J0030+0451 exhibits a periodic mode-changing phenomenon, which has never reported in MSPs before. 

We also note that the maximum energy of MP exceeds 10$\langle E\rangle$, which suggests the existence of giant pulses.  
The profile of the brightest pulse is shown in  Figure~\ref{0030sp} with a W50 width of 58\,$\mu$s, much narrower than the average pulse of 577\,$\mu$s. 
The energy and peak intensity of this pulse are about 11 and 70 times greater than the average pulse, respectively. 
PSR J0030+0451 might be a potential giant pulse emitter. 
However, due to the limitations of our observation with an 8.192 $\mu$s interval, it is insufficient to identify typical giant pulses with nanosecond durations. Further observations with a higher sampling interval are required.

The measured $\sigma_{\rm J}({\rm h})$ at different frequencies for previously reported results are shown in Figure~\ref{fj}, where there is a tendency that the jitter noise increases with observing frequency. 
Our measured $\sigma_{\rm J}({\rm h})$ is $43.4\pm0.1\,$ns at 1250\,MHz, which is smaller than the results of 60\,ns at 1284\,MHz~\citep{Parthasarathy2021}, 153\,ns at 1400\,MHz~\citep{Lam2016}, and $61.1^{+3.5}_{-3.5}$\,ns at 1500\,MHz~\citep{Lam2019}. Therefore, it is reasonable that our measured jitter noise at 1250\,MHz is smaller than the reported values at higher observing frequencies.

\subsubsection{PSR J0613$-$0200}

PSR J0613$-$0200 exhibits three components, namely C1,C2 and C3 (Figure~\ref{gray}). 
Among these components, C2 exhibits greater flux density compared to C1 and C3.
We performed LRFS analysis for this pulsar, and the results are presented in Figure~\ref{modulation}. We detected periodic modulations across the C1 and C2 components, with modulation periods of $6\pm1\,P$ and $72.9\pm0.4\,P$, respectively.

The measured $\sigma_{\rm J}({\rm h})$ at different frequencies for previously reported results are shown in Figure~\ref{fj}. Similar to the case of PSR J0030+0451, there is also a tendency for the jitter noise to increase with observing frequency for PSR J0613$-$0200. Our measured $\sigma_{\rm J}({\rm h})$ is $34.4\pm0.2$\,ns at 1250\,MHz, which roughly agrees with the results of $<400$\,ns at 1400 MHz~\citep{Shannon2014}, of $<44$\,ns at 1400 MHz~\citep{Lam2016}, but is much smaller than the result of $133^{+8}_{-8}$\,ns at 1500\,MHz~\citep{Lam2019}.

\subsubsection{PSR J0636+5128}

PSR J0636+5128 exhibits a single component in its pulse profile (Figure~\ref{gray}).
The pulsar exhibits complicated periodical modulations (Figure~\ref{modulation}). A modulation period of $84.3\pm0.9\,P$ occurs across the entire profile, while modulation periods of $2.20\pm0.01\,P$ and $2.26\pm0.01\,P$ occur only in the leading and trailing parts of the profile, respectively.

The measured $\sigma_{\rm J}({\rm h})$ at different frequencies for previously reported results are shown in Figure~\ref{fj}. 
Our measured $\sigma_{\rm J}({\rm h})$ is $29.6\pm0.4\,$ns at 1250\,MHz, which is smaller than the results of $62^{+16}_{-22}$\,ns at 1500\,MHz and $47^{+13}_{-11}$\,ns at 820\,MHz~\citep{Lam2019}, but with much smaller uncertainties.

\subsubsection{PSR J0751+1807}

PSR J0751+1807 exhibits two components, C1 and C2, with C2 being stronger (Figure~\ref{gray}). 
The bright single-pulses occur across both the C1 and C2 components (Figure~\ref{gray}). We performed LRFS analysis and found periodical modulations in both components (Figure~\ref{modulation}). The modulation period of C1 is $128.0\pm0.1\,P$, while the modulation period of the C2 is $6.3\pm0.1\,P$. The correlation coefficient between the energy of the two components is 0.35, indicating no significant correlation between them.
We report new jitter measurement of this pulsar with $\sigma_{\rm J}({\rm h})=32.5\pm0.5$\,ns.

\subsubsection{PSR J1012+5307}

PSR J1012+5307 exhibits a main pulse and an interpulse in its pulse profile (Figure~\ref{gray}). 
The pulsar presents complex modulations in both the main pulse and interpulse~\citep{Edwards2003}. Our highly sensitive observation allows for a more detailed study of the modulations (Figure~\ref{gray}). By calculating the LRFS, we found that the two components of the mean pulse (C1 and C2) exhibit different modulation periods, with modulation periods of 7.1$\pm$0.1$P$ and 7.6$\pm$0.1$P$, respectively. For the interpulse, periodical modulation is observed only in the first and second components (C3 and C4), with a similar modulation period of $11.1\pm0.1\,P$.

The measured $\sigma_{\rm J}({\rm h})$ at different frequencies for previously reported results are shown in Figure~\ref{fj},  showing a tendency for the jitter noise to increase with observing frequency. 
Our measured $\sigma_{\rm J}({\rm h})$ is $27.0\pm0.2\,$ns at 1250\,MHz, which roughly agrees with the result of $<103$\,ns at 1400\,MHz~\citep{Lam2016}, but is smaller than the result of $67\pm6$\,ns at 1500\,MHz~\citep{Lam2019}.

\subsubsection{PSR J1643$-$1224}

PSR J1643$-$1224 exhibits a single component in its pulse profile (Figure~\ref{gray}). 
The pulsar exhibits periodical modulation with a period of $46\pm1\,P$ occurring across the leading part of the profile, as determined by LRFS analysis (Figure~\ref{modulation}).

The measured $\sigma_{\rm J}({\rm h})$ at different frequencies from previously reported results are presented in Figure~\ref{fj}, showing a tendency for the jitter noise to increase with observing frequency. 
Our measured $\sigma_{\rm J}({\rm h})$ is $40.8\pm0.3$\,ns at 1250\,MHz, which roughly agrees with the result of $<60$\,ns at 1284\,MHz~\citep{Parthasarathy2021} and $<500$\,ns at 1400\,MHz~\citep{Shannon2014}, but is smaller than the results of $155$\,ns at\,1400 MHz~\citep{Lam2016} and $120^{+8}_{-10}$\,ns at 1500\,MHz~\citep{Lam2019}.

\subsubsection{PSR J1713+0747}

PSR J1713+0747 is a bright MSP. The profile of the pulsar changed between April 15, 2021, and April 17, 2021~\citep{Xu2021}, and shows a slow recovery continuing for the last several years~\citep{Jennings2022}. Our observation was on August 20, 2021, during which the profile had not completely recovered yet.
PSR J1713+0747 exhibits multiple components, but it is dominated by a central component (C1) in its pulse profile (Figure~\ref{gray}). We identified a periodical modulation in our observation with a period of $6.2\pm0.1\,P$ (Figure~\ref{modulation}).
Considering the diffuse modulation, we think that our result is roughly consistent with the previous result of $6.9\pm0.1\,P$ reported by \citet{Liu2016}.

The measured $\sigma_{\rm J}({\rm h})$ at different frequencies for previously reported results are shown in Figure~\ref{fj}, where there is a tendency that the jitter noise decreases with observing frequency. Our measured $\sigma_{\rm J}({\rm h})$ is $24.9\pm0.3$\,ns at 1250\,MHz, which is smaller than the results of $35.0\pm0.8$\,ns at 1400\,MHz~\citep{Shannon2014}, $28/36$\,ns at 1400\,MHz~\citep{Lam2016}, and $29.3^{+3.2}_{-0.9}$\,ns at 1500\,MHz~\citep{Lam2019}. Note that the profile in our observation has not recovered yet, which may  cause the change in the level of jitter noise. For pulsars exhibiting mode changing phenomena, the levels of jitter noise in different modes can vary, as seen in PSR J1909$-$3744~\citep{Miles2022}.

\subsubsection{PSR J1744$-$1134}

PSR J1744$-$1134 exhibits a main pulse and a weak interpulse in its pulse profile (Figure~\ref{gray}). 
The measured $\sigma_{\rm J}({\rm h})$ at different frequencies for previously reported results are shown in Figure~\ref{fj}, indicating a pattern where the jitter noise decreases and then increases with observing frequency.
Our measured $\sigma_{\rm J}({\rm h})$ is $29.4\pm0.2$\,ns at 1250\,MHz, which agrees with the results of $31$\,ns at 1400\,MHz~\citep{Lam2016} and $30\pm6$\,ns at 1284\,MHz~\citep{Parthasarathy2021}, but is smaller than the results of $37.8\pm0.8$\,ns at 1400\,MHz~\citep{Shannon2014} and $46.5\pm1.3$\,ns at 1500\,MHz~\citep{Lam2019}.

\subsubsection{PSR J1911+1347}

PSR J1911+1347 exhibits a relatively bright pulse component (C1) surrounded by broader emission in its pulse profile (Figure~\ref{gray}). The narrow bright pulses primarily occur in the C1 component. 
The measured $\sigma_{\rm J}({\rm h})$ at different frequencies for previously reported results are shown in Figure~\ref{fj}, showing an increase in jitter noise with observing frequencies. Our measured $\sigma_{\rm J}({\rm h})$ is $27.8\pm0.2$\,ns at 1250\,MHz, which is smaller than the result of $43.5^{+2.3}_{-2.3}$\,ns at 1500\,MHz~\citep{Lam2019}.

\subsubsection{PSR J1918$-$0642}

PSR J1918$-$0642 exhibits a main pulse and an interpulse in its pulse profile. 
The main pulse exhibits a relatively bright pulse component (C1) with broader emission flanking it, and the narrow bright pulse only occurs in the C1 (Figure~\ref{gray}). 
Previous work by \citet{Edwards2003} suggested that PSR J1918$-$0642 may exhibit quasi-periodic modulations with periods ranging from 2 to 4\,$P$, but the limited sensitivity of their study made their results uncertain.
In our analysis, we detected periodical modulation in this pulsar with a period of $3.3\pm0.1\,P$ through the calculation of the LRFS (Figure~\ref{modulation}). However, the modulation only occurs across the C1 of the main pulse.

The measured $\sigma_{\rm J}({\rm h})$ at different frequencies for previously reported results are shown in Figure~\ref{fj}, with the jitter noise increasing as the observing frequency increases. Our measured $\sigma_{\rm J}({\rm h})$ is $55.7\pm0.5$\,ns at 1250\,MHz, which agrees with the results of $<101$\,ns at 1400\,MHz~\citep{Lam2016}, $56^{+9}_{-13}$\,ns at 1500\,MHz~\citep{Lam2019}, and $<$55\,ns at 1284\,MHz~\citep{Parthasarathy2021}.

\subsubsection{PSR J1944+0907}

PSR J1944+0907 exhibits a wide profile with about four components (C1, C2, C3, and C4) in its pulse profile (Figure~\ref{gray}). The bright narrow single-pulse is primarily observed in the C1.
We identified periodical modulation that occurs across the C1, C2, and C3 components with the same modulation period of $128\pm2\,P$ by calculating the LRFS  (Figure~\ref{modulation}).

The measured $\sigma_{\rm J}({\rm h})$ at different frequencies for previously reported results are shown in Figure~\ref{fj}, with the jitter noise increasing as the observing frequency increases. Our measured $\sigma_{\rm J}({\rm h})$ is $174\pm1$\,ns at 1250\,MHz, which is smaller than the results of $196$\,ns at 1400\,MHz~\citep{Lam2016} and $311^{+6}_{-7}$\,ns at 1500\,MHz~\citep{Lam2019}.

\subsubsection{PSR J2145$-$0750}

PSR J2145$-$0750 is a bright millisecond pulsar with multiple components in its pulse profile (Figure~\ref{gray}). 
We observed periodical modulation occurring across the C1 component with a modulation period of $3.9\pm0.1\,P$ (Figure~\ref{modulation}), which is consistent with the previous result of $\sim2-4\,P$~\citep{Edwards2003}.

The measured $\sigma_{\rm J}({\rm h})$ at different frequencies for previously reported results are shown in Figure~\ref{fj}, where no obvious tendency of increase/decrease with observing frequency is detected. Our measured $\sigma_{\rm J}({\rm h})$ is $168\pm1$\,ns at 1250\,MHz. This value agrees with the result of $173^{+3}_{-4}$\,ns at 1500\,MHz~\citep{Lam2019}, but is larger than the result of $85$\,ns at 1400\,MHz~\citep{Lam2016}, and smaller than the result of $192\pm6$\,ns at 1400\,MHz~\citep{Shannon2014} and $200\pm20$\,ns at 1284\,MHz~\citep{Parthasarathy2021}.

\subsection{Timing a subselection of single-pulses}

\begin{figure*}
\centering
\includegraphics[width=40mm]{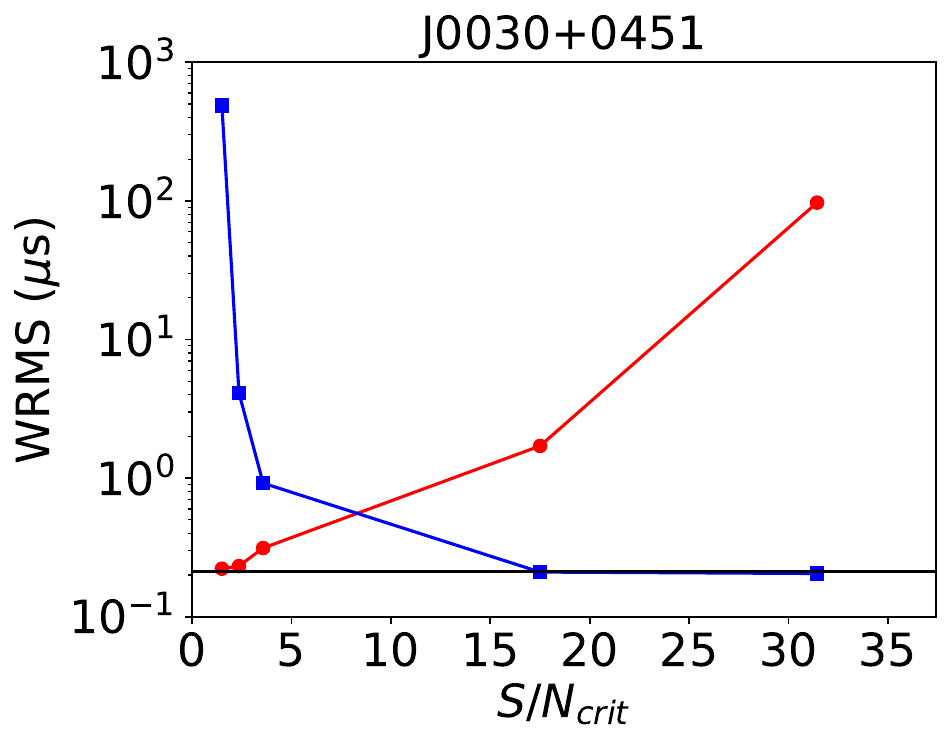}
\includegraphics[width=40mm]{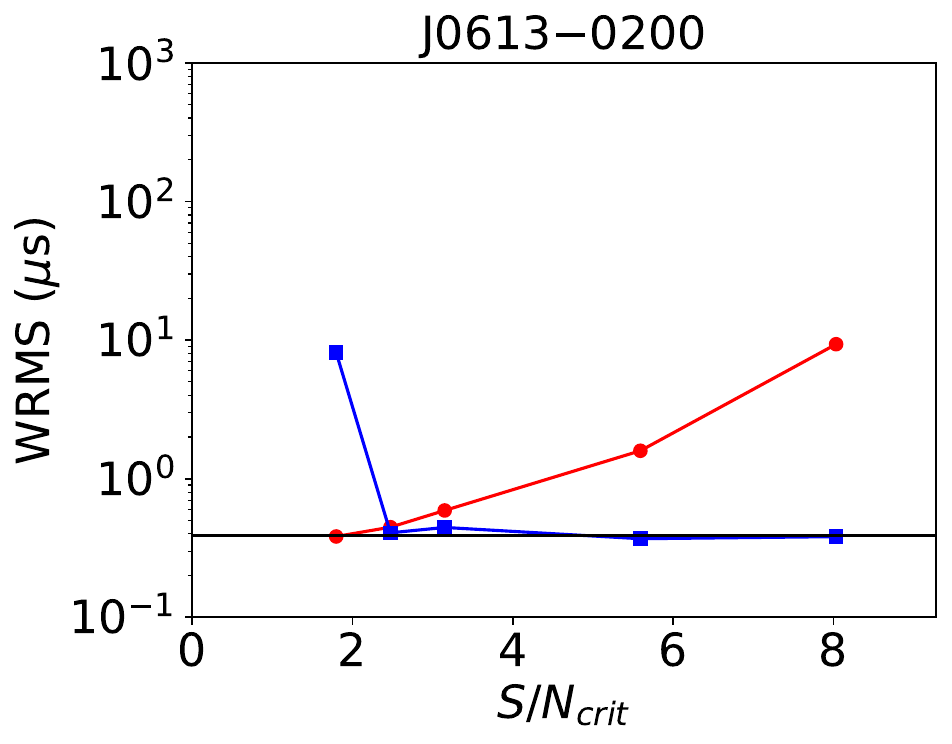}
\includegraphics[width=40mm]{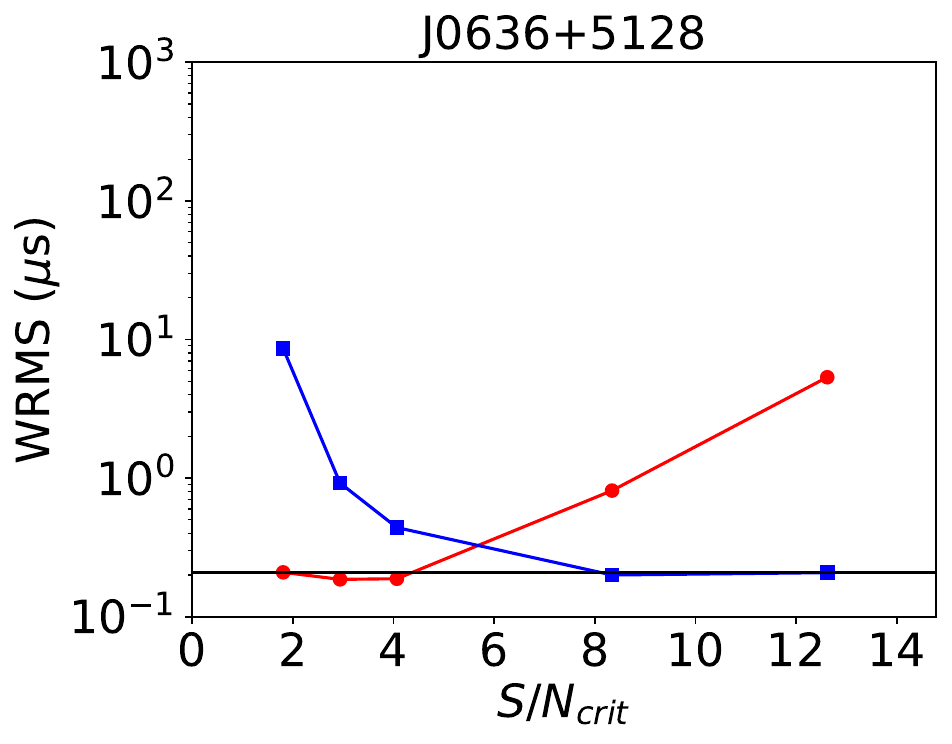}
\includegraphics[width=40mm]{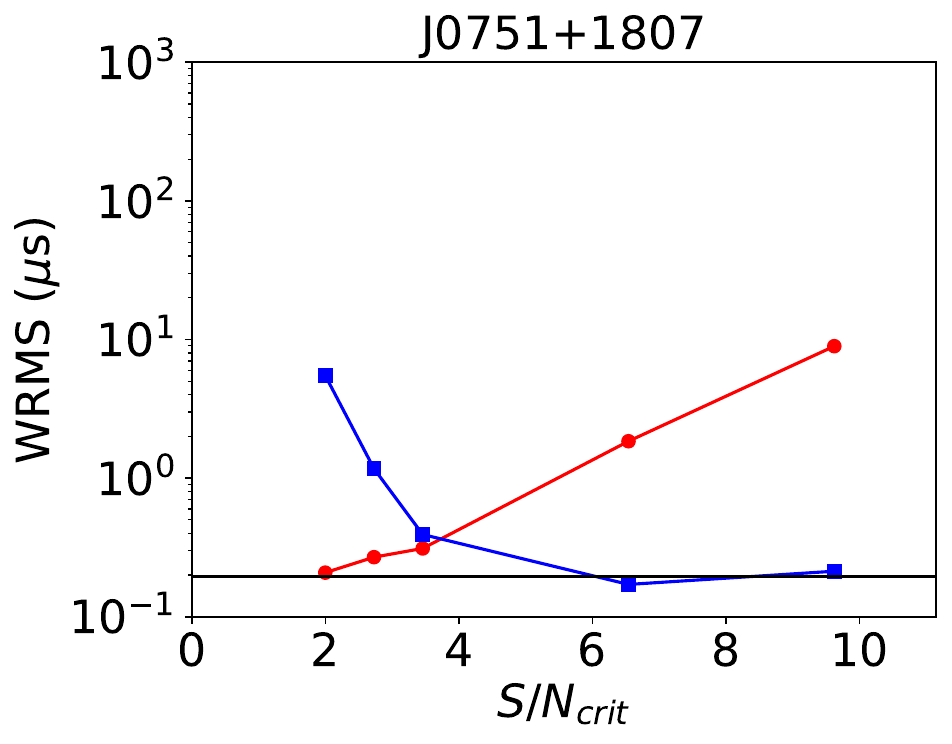}
\includegraphics[width=40mm]{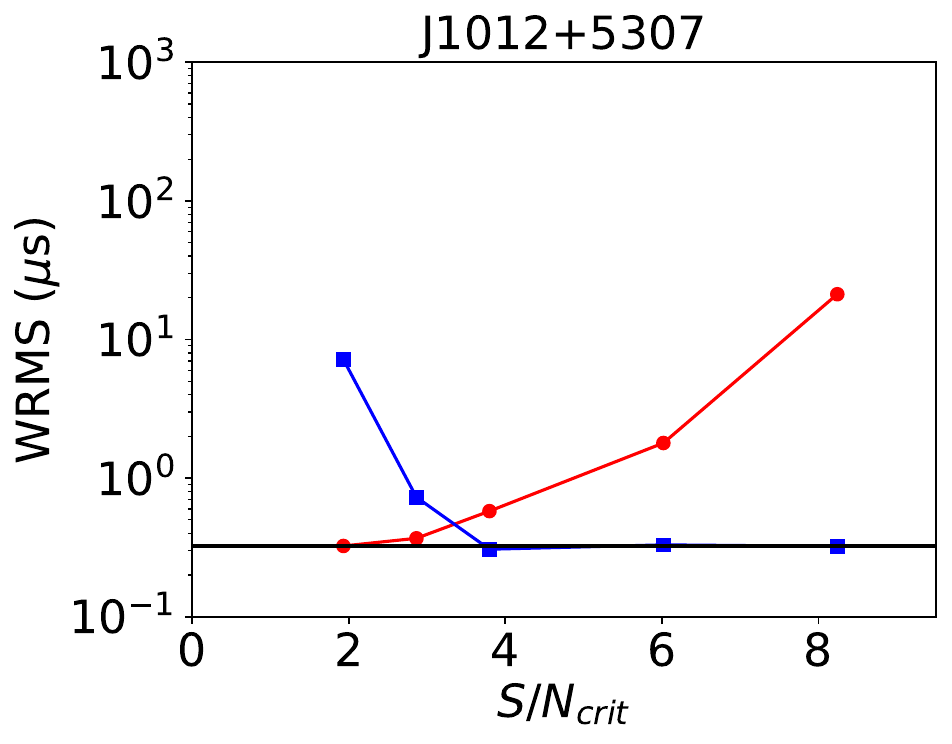}
\includegraphics[width=40mm]{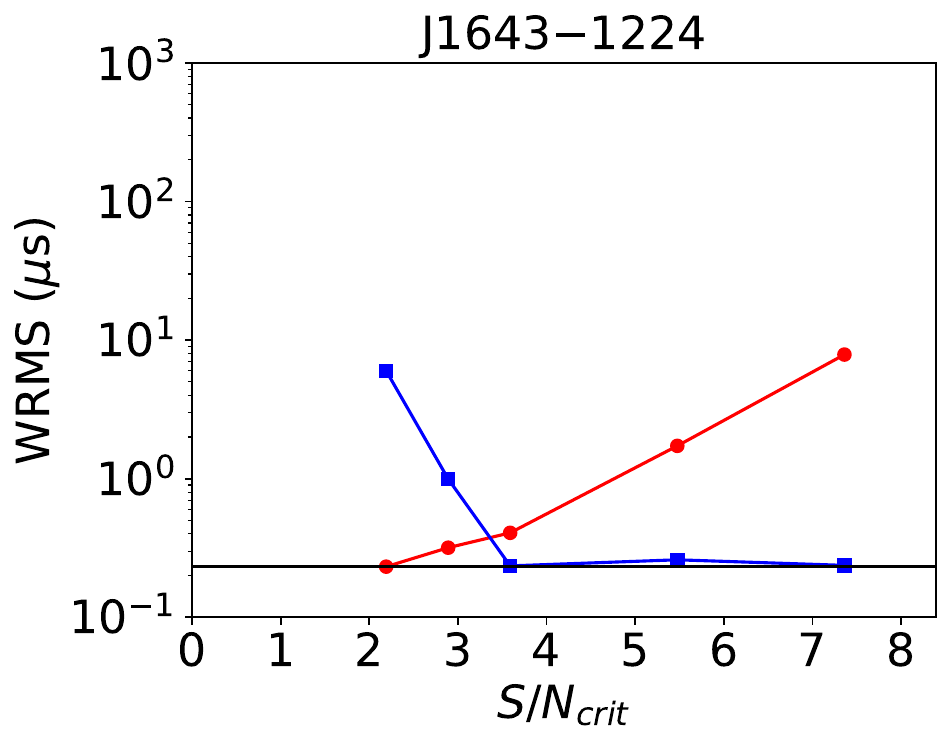}
\includegraphics[width=40mm]{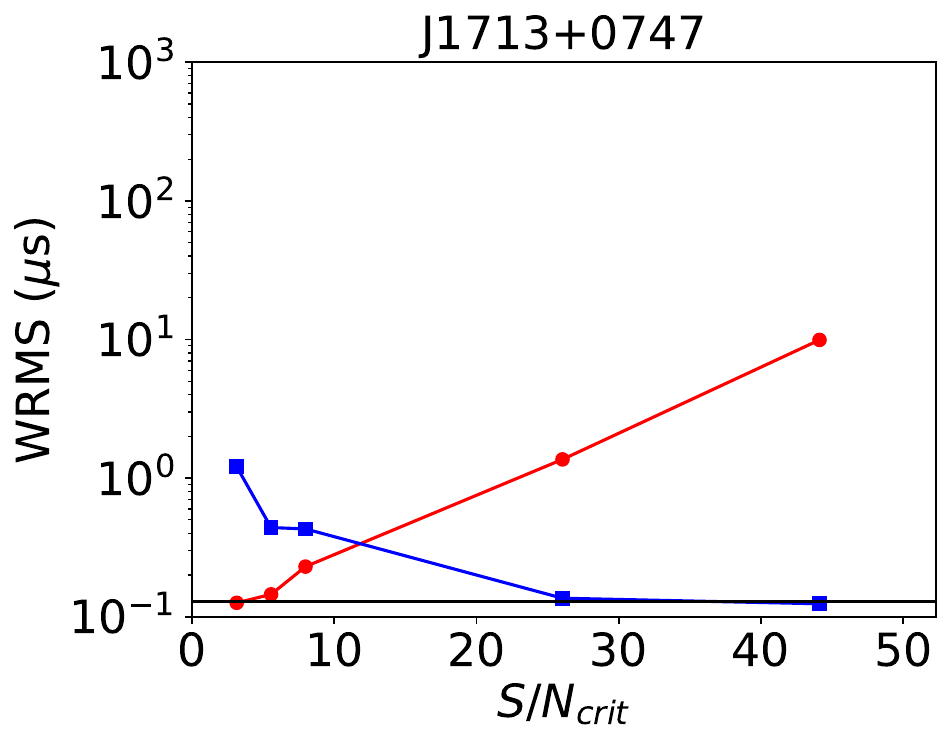}
\includegraphics[width=40mm]{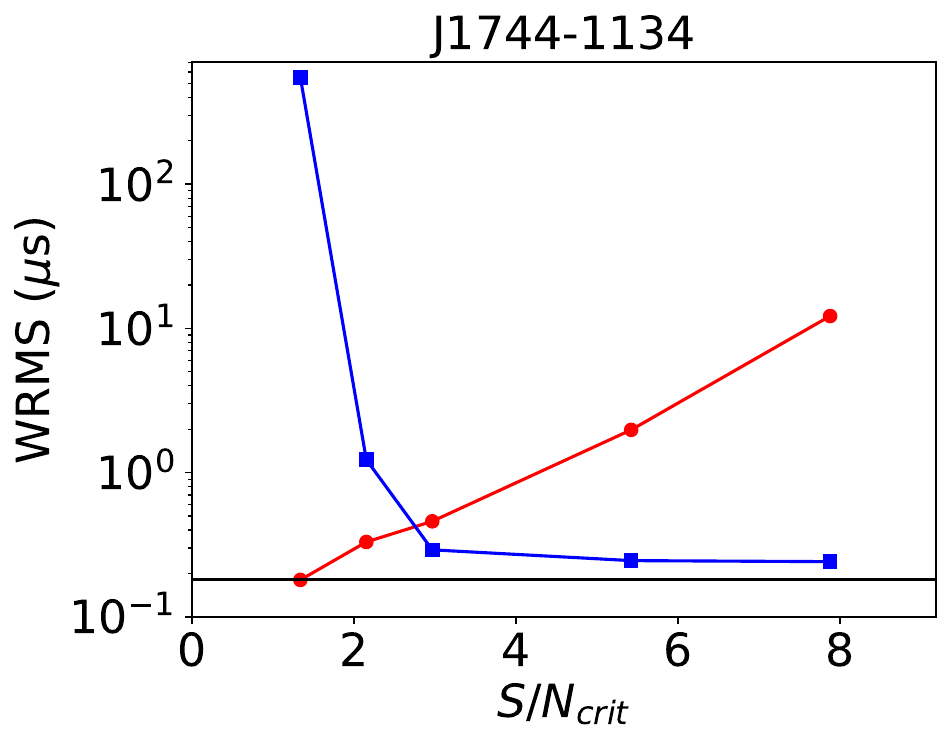}
\includegraphics[width=40mm]{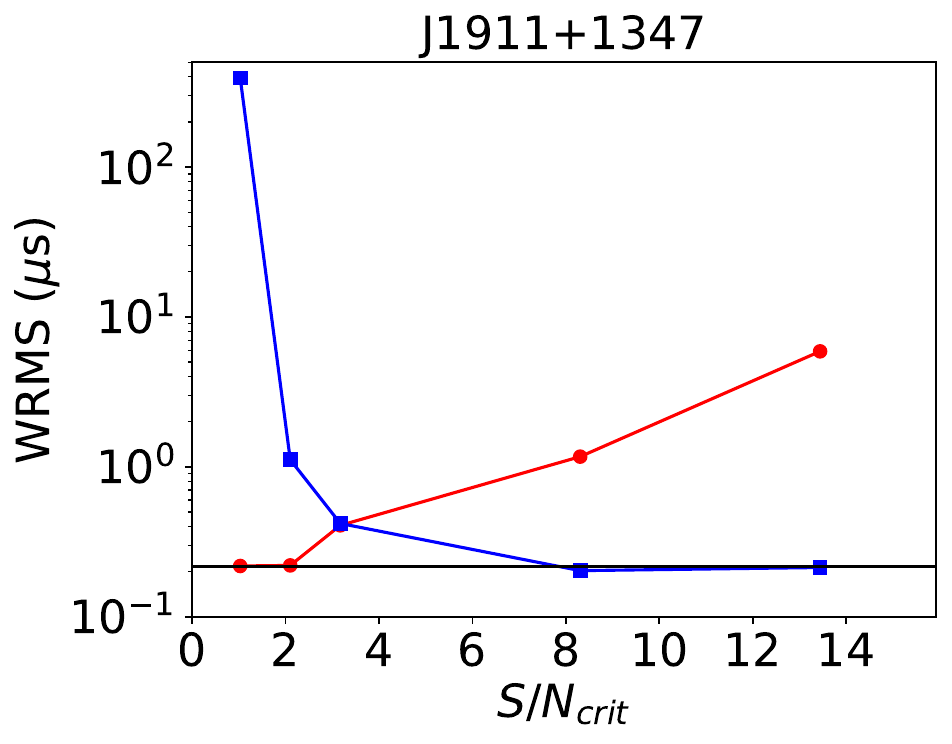}
\includegraphics[width=40mm]{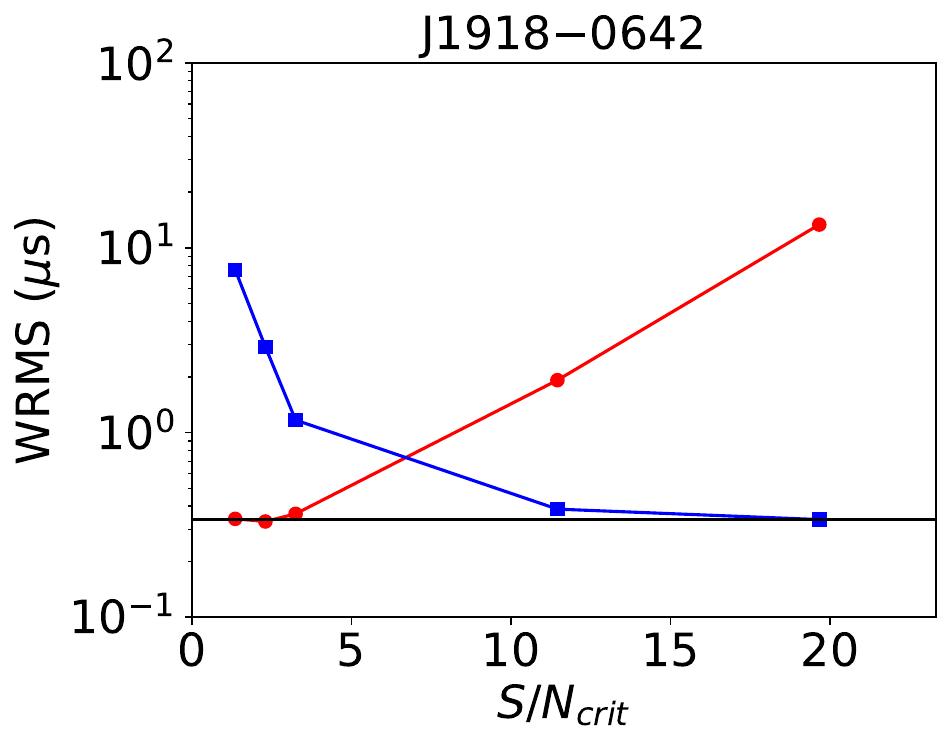}
\includegraphics[width=40mm]{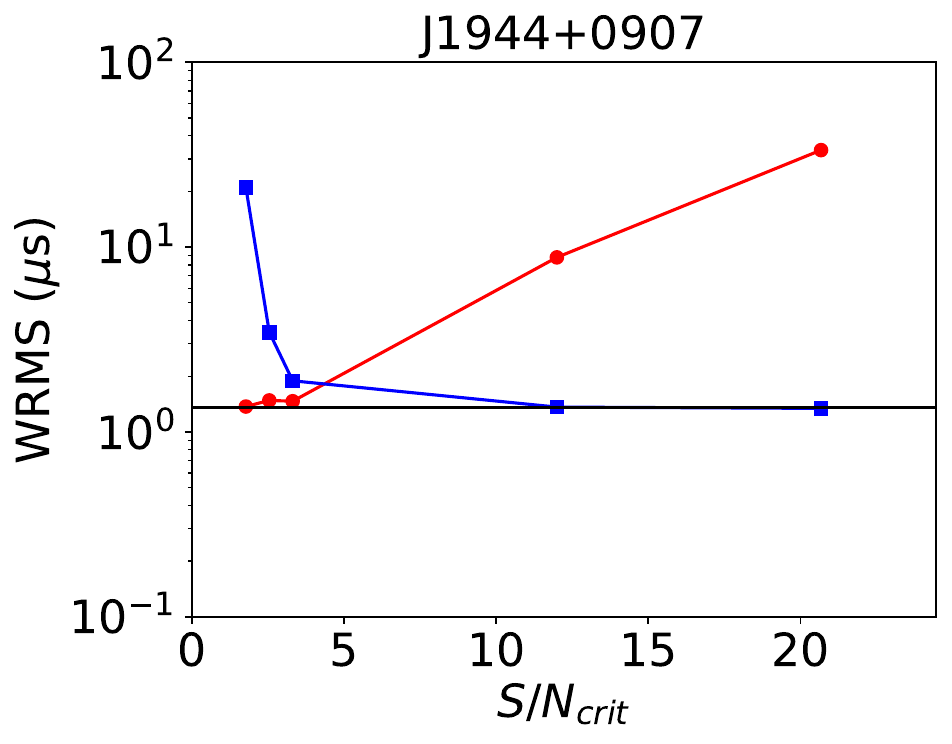}
\includegraphics[width=40mm]{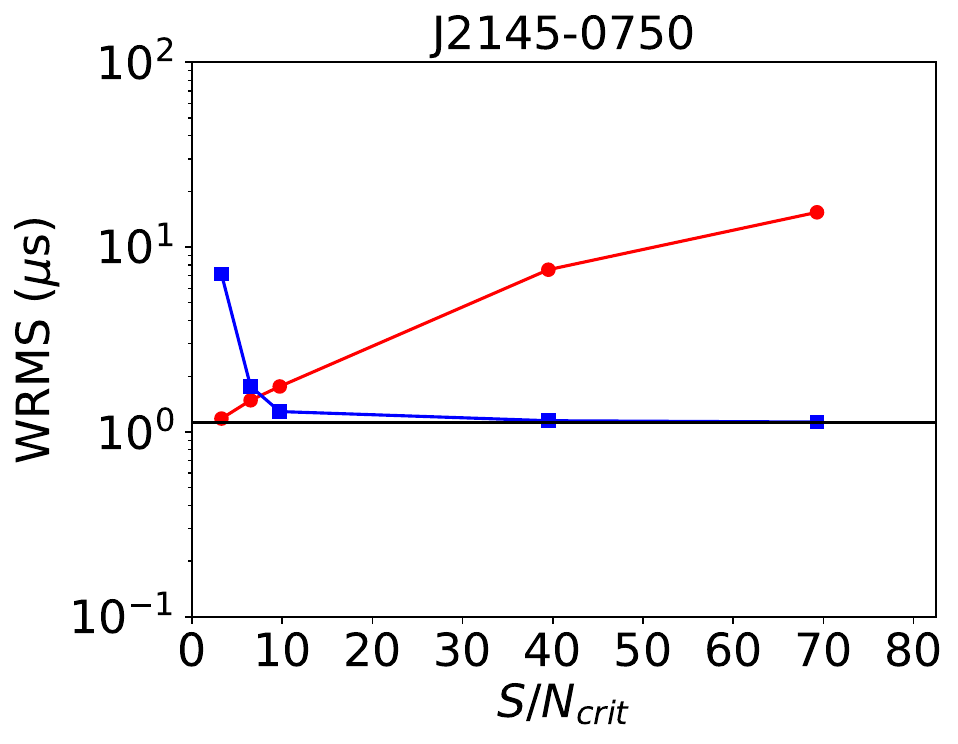}

\caption{Wrms of the bright (red circles) and weak (blue squares) pulses at different $S/N_{\rm crit}$ for each pulsar. The black line is for the Wrms in 60\,s subintegrations using all the single-pulses. Note that the y-axes is plotted on the same scale.}
\label{sn}
\end{figure*}

The variability of single-pulses introduces jitter noise, which can have a significant impact on highly sensitive observations. 
It may be possible to improve the timing precision by selecting a sub-set of single pulses. 
\citet{Miles2022} conducted timing analysis on PSR J1909$-$3744 which shows mode changing phenomenon, and found that timing precision improved by approximately 10\% by only timing the strong mode.
 \citet{Oslowski2014} studied the rms of the timing residual of PSR J0437$-$4715 for average profiles formed by integrating for one minute but rejecting the brightest single pulses with different S/N. They found that the rms of the timing residual decreases with the increasing S/N, but no significantly improvement is achievable.
Following the method of  \citet{Oslowski2014}, we divided the observation into several segments, each lasting 60\,s. Within each segment, we separately summed all the pulses, as well as the bright and weak pulses. 
We then formed three pulse templates based on the average profiles of the bright, weak, and all of the pulses from the entire observation using the {\sc PAAS}. 
Pulse ToAs were obtained by cross-correlating the summed profiles of the segments with the corresponding standard template, and timing residuals were calculated using the {\sc TEMPO2} software (a similar method was demonstrated in \citealt{Wang2020}). We considered timing precision improvement achievable if the weighted rms (Wrms) of the timing residuals obtained using a sub-selection of pulses was lower than that obtained using all of the pulses.

We note that there are some bright pulses with narrow pulse widths pulsars (Figure~\ref{gray}). Therefore, we used $S/N_{\rm peak}$ to define bright pulses. We defined a threshold value ($S/N_{\rm  crit}$) to classify pulses as either bright or weak. If the $S/N_{\rm peak}$ of a single-pulse is greater than $S/N_{\rm crit}$, it is classified as a bright pulse; otherwise, it is classified as a weak pulse. For each pulsar, we divided the single-pulses into five groups based on $S/N_{\rm crit}$ (Table~\ref{crit}). Here, $S/N_{\rm mean}$ represents the mean $S/N_{\rm peak}$, while $S/N_{\rm max}$ and $S/N_{\rm min}$ are the maximum and minimum $S/N_{\rm peak}$ values, respectively. We then calculated the Wrms of the timing residuals for the bright and weak pulses at these five $S/N_{\rm crit}$ values for each pulsar (Figure~\ref{sn}). 
As the number of bright pulses decreases with increasing $S/N_{\rm crit}$, it is expected that the Wrms of the bright pulses increases (the red lines in Figure~\ref{sn}), while that of the weak pulses decreases (the blue lines in Figure~\ref{sn}). 
 However, when comparing the timing precision achieved using a sub-set of pulses with a specific range of $S/N_{\rm peak}$ to the timing precision achieved using all of the single-pulses (the black lines in Figure~\ref{sn}), we did not observe a improvement in timing precision for the pulsars in our sample, just like the case of PSR J0437$-$4715~\citep{Oslowski2014}.

\begin{table}  
\centering
\caption{The threshold value ($S/N_{\rm crit}$) that was used to classify the bright/weak pulses.} 
\label{crit}
\begin{tabular}{cc}
\hline
Groups &   $S/N_{\rm crit}$   
\\
\hline
1 &  $S/N_{\rm mean}-2/3(S/N_{\rm mean}-S/N_{\rm min})$
\\
2 & $S/N_{\rm mean}-1/3(S/N_{\rm mean}-S/N_{\rm min})$
\\
3 &  $S/N_{\rm mean}$ 
\\
4 & $S/N_{\rm mean}+1/3(S/N_{\rm max}-S/N_{\rm main})$,
\\
5 & $S/N_{\rm mean}+2/3(S/N_{\rm max}-S/N_{\rm main})$
\\
\hline
 \end{tabular}
\end{table}

\section{Discussion And Conclusions}\label{sec: conclusions}

We presented measurements of jitter noise, radiometer noise, and scintillation noise for 12 millisecond pulsars that are part of the IPTA sample, using FAST. 
In our analysis, we determined that the mean levels of jitter noise, radiometer noise, and scintillation noise in a one-hour-long duration are approximately 58\,ns, 33\,ns, and 4\,ns, respectively. 
This indicates that the contribution of scintillation noise is likely negligible, being roughly one-tenth the level of the jitter/radiometer noise, consistent with previous published results (e.g., \citet{Lam2016}).
Our results suggest that jitter noise significantly contributes to white noise, emphasizing its notable impact on PTA observations with FAST.

Jitter noise is known to exhibit frequency dependence~\citep{Lam2019}. When considering previously published results on jitter noise, we observe a diverse trend: while the jitter noise for some pulsars increases with observing frequency, it decreases for others. \citet{Lam2019} conducted a comprehensive analysis of the frequency dependence of jitter noises in a large sample of MSPs, revealing a wide range of indexes in a power-law model, spanning from $-3.2$ to 9. MSPs typically demonstrate simple power-law spectra with a mean index of approximately $-1.67$~\citep{Dai2015}, resulting in radiometer noise increasing with observing frequency.
For pulsars, the interplay between jitter noise and radiometer noise at varying observing frequencies becomes crucial. If the jitter noise increases with frequency, exhibiting a spectrum steeper than the radiometer noise, it dominates at higher frequencies; otherwise, radiometer noise prevails. Optimizing observations can be achieved through the use of multiple telescopes with distinct observing frequencies. For instance, higher frequency observations for specific pulsars might significantly imporve arrival-time precision~\citep{Lam2018}. The Qitai Radio Telescope (QTT), currently under construction, will enable high-sensitivity observations spanning from 150 MHz up to 115 GHz~\citep{wxm+23}. The synergy between FAST and QTT will promise a substantial contribution to the capabilities of IPTA.

Our results indicate the substantial impact of jitter noise on PTA with FAST. The investigation of single pulses represents a potential avenue for improving timing precision in MSPs. For instance, \citet{Miles2022} conducted timing analysis on PSR J1909$-$3744, which exhibits mode-changing phenomena. They observed an approximately 10\% improvement in timing precision by exclusively considering the strong mode. In our sample, where some pulsars exhibit bright and narrow single pulses or periodic intensity modulations, we explored the potential for improving timing precision by selecting specific subsets of single pulses. However, our analysis revealed that selecting a subset based on a specific range of $S/N_{\rm peak}$ did not improve achievable timing precision.

To improve pulsar timing precision, considering information about polarized radiation has been proposed~\citep{Straten06,Oslowski13}. Accordingly, we conducted matrix template matching using all four Stokes parameters (\citealt{Straten06}) for each pulsar to generate pulse ToAs, and measured radiometer and jitter noise. We successfully detected jitter noise in 8 pulsars. Notably, we observed a decrease in jitter noise ranging from 6.7\% to 39.6\%, while radiometer noise increased by 6.1\% to 38.6\%. The total white noise also decreased for 7 out of the 8 pulsars in our sample, ranging from 0.19\% to 13.4\%. 
Therefore, matrix template matching using all four Stokes parameters is a valuable method for reducing jitter noise in pulsars.
In future work, we plan to generate pulse ToAs by considering all four Stokes parameters on a large dataset of MSPs using FAST, and present the noise analysis of them with the expectation of further improving pulsar timing precision.

The highly sensitive and rapidly sampled observations conducted with FAST have provided us with the opportunity to study single pulses in a larger sample of MSPs. Utilizing LRFS analysis, we identified periodical intensity modulations in 10 pulsars: PSR J0030+0451, PSR J0613$-$0200, PSR J0636+5128, PSR J0751+1807, PSR J1012+5307, PSR J1643$-$1224, PSR J1713+0747, PSR J1918$-$0642, PSR J1944+0907, and PSR J2145$-$0750. Notably, 6 of these pulsars were reported for the first time. Our findings for PSR J1918$-$0642, PSR J1012+5307, PSR J2145$-$0750, and PSR J1713+0747 are consistent with previous results from \citet{Edwards2003} and \citet{Liu2016}.
PSR J0030+0451, PSR J0636+5128, and PSR J1944+0907 exhibit coherent periodical modulations, whereas the remaining 7 pulsars show diffuse periodical modulations. The modulation periods for these MSPs range from approximately 6 to 665 ms (2 to 128 P), which is notably shorter than those observed in normal pulsars, typically ranging from a few seconds to several minutes~\citep{Basu2020}. The origin of periodic modulations in normal pulsars and MSPs remains unknown. In normal pulsars, these modulations are attributed to global changes in currents within the pulsar magnetosphere~\citep{Latham2012} and are generally observed across the entire pulse profile~\citep{Basu2020}. Some normal pulsars exhibit phase-locked modulations of the main pulse and interpulse~\citep{Weltevrede2007}.
However, in MSPs, periodic modulations may occur only across a portion of the profile. We did not observe phase-locked phenomena in MSPs with interpulse emission. For example, in PSR J1012+5307, the modulation periods of the mean pulse and interpulse are different. In PSR J0030+0451, periodical modulation is only detected in the mean pulse. The distinct modulation characteristics between MSPs and normal pulsars may be attributed to the presence of multiple emission regions in the magnetospheres of MSPs, with different pulse components originating from distinct locations (e.g., \citealt{Dyks2010}).

In particular, we identified brighter, narrower single pulses in PSR J0030+0451, with energies exceeding 10 times the mean pulse energy. This strongly suggests the presence of giant pulses. Additionally, we observed a periodic mode changing phenomenon in PSR J0030+0451, a behavior not previously reported in MSPs.
Until now, mode changing phenomena have only been detected in two MSPs, namely PSR B1957+20~\citep{Mahajan2018} and PSR J1909$-$3744~\citep{Miles2022}. However, the mode cycles observed in PSR B1957+20 and PSR J1909$-$3744 are not periodic. Further observations with FAST, encompassing a larger sample of MSPs, may unveil more instances of diverse emission behaviors, such as mode changing, nulling, or giant pulses. This expanded dataset could provide valuable constraints on the emission properties of MSPs.

\section*{Acknowledgments}

This is work is supported by the National Natural Science Foundation of China (No. 12288102, No. 12203092, No. 12041304),  the Major Science and Technology Program of Xinjiang Uygur Autonomous Region (No. 2022A03013-3), the National SKA Program of China (No. 2020SKA0120100), the National Key Research and Development Program of China (No. 2022YFC2205202, No. 2021YFC2203502), the Natural Science Foundation of Xinjiang Uygur Autonomous Region (No. 2022D01B71, No. 2022D01D85), the Tianshan talent Program and the Special Research Assistant Program of CAS, the CAS Project for Young Scientists in Basic Research (No. YSBR-063), the Key Research Project of Zhejiang Laboratory (No. 2021PE0AC03). the Zhejiang Provincial Natural Science Foundation of China under Grant (No. LY23A030001). 
This work made use of the data from the Five-hundred-meter Aperture Spherical radio Telescope, which is a Chinese national megascience facility, operated by National Astronomical Observatories, Chinese Academy of Sciences. The research is partly supported by the Operation, Maintenance and Upgrading Fund for Astronomical Telescopes and Facility Instruments, budgeted from the Ministry of Finance of China (MOF) and administrated by the Chinese Academy of Sciences (CAS).

\software{DSPSR \citep{Straten2011}, PSRCHIVE \citep{Hotan2004}, TEMPO2 \citep{Hobbs2006}, and PSRSALSA \citep{Weltevrede2016}}

\bibliography{sample63}{}
\bibliographystyle{aasjournal}

\end{document}